\documentclass[9pt,twocolumn]{ieeetran}

\usepackage{amstex}
\usepackage{epsfig}
\usepackage{amssymb}
\usepackage{latexsym}
\usepackage{graphics}
\usepackage{supertab}

\input epsf
\input{psfig.sty}

\newlength{\pgmtab}          
\begin{document}

\title{\bf\Large Systematic Testing of Multicast Routing Protocols: Analysis
of Forward and Backward Search Techniques}

\author{Ahmed Helmy, Deborah Estrin, Sandeep Gupta\\ 
University of Southern California, Los Angeles, CA 90089\\ 
email: helmy@@ceng.usc.edu, estrin@@usc.edu, sandeep@@boole.usc.edu}

\maketitle

\begin{abstract}



The recent growth of the Internet and its increased heterogeneity
have increased the complexity of network protocol design and
testing. 
In addition, the advent of multipoint (multicast-based) applications
has introduced new challenges that are qualitatively different in
nature than the traditional point-to-point protocols. 
Multipoint applications typically involve a group of participants
simultaneously, and hence are inherently more complex. As more
multipoint protocols are coming to life, the need for a systematic 
method to study and evaluate such protocols is becoming
more apparent. Such method aims to expedite the protocol development
cycle and improve protocol robustness and performance.

In this paper, we present a new methodology for developing systematic and
automatic test generation algorithms for multipoint protocols. 
These algorithms attempt to synthesize network topologies
and sequences of events that stress the protocol's correctness or
performance. This problem can be viewed as a domain-specific search
problem that suffers from the state space explosion problem. One goal
of this work is to circumvent the state space explosion problem
utilizing knowledge of network and fault modeling, and multipoint
protocols.
The two approaches investigated in this study are based on
forward and backward search techniques.
We use an 
extended finite state machine (FSM) model of the protocol. 
The first algorithm uses forward search to perform reduced
reachability analysis. Using domain-specific information for multicast
routing over LANs, the algorithm complexity is reduced from
exponential to polynomial in the number of routers. This approach,
however, does not fully automate topology synthesis. The second
algorithm, the fault-oriented test generation, uses backward search for topology
synthesis and uses backtracking to generate event sequences instead of searching
forward from initial states.

Using these algorithms, we have conducted studies for correctness of  
the multicast routing protocol PIM. We propose to extend these algorithms 
to study end-to-end multipoint protocols using a virtual LAN 
that represents delays of the underlying multicast distribution tree.
\end{abstract}

\section{Introduction}
\label{ch-intro}

Network protocols are becoming more complex with the 
exponential growth of the Internet, and the introduction of new services 
at the network, transport and application levels. 
In particular, the advent of IP multicast and the MBone enabled 
applications ranging from multi-player games to distance learning and 
teleconferencing, among others.
To date, little effort has been exerted to formulate systematic methods 
and tools that aid in the design and characterization of these protocols.

In addition, researchers are observing new and obscure, yet all too 
frequent, failure modes over the
internets~\cite{paxon}~\cite{paxon2}. Such 
failures
are becoming more frequent, mainly due to the increased heterogeneity of 
technologies, interconnects and configuration of various network components.
Due to the synergy and interaction between different
network protocols and components, errors at one layer may lead to
failures at other layers of the protocol stack. Furthermore,
degraded performance of low level network protocols may have
ripple effects on end-to-end protocols and applications.

Network protocol errors are often detected by application failure or 
performance degradation. Such errors are hardest to diagnose when the 
behavior is unexpected or unfamiliar.
Even if a protocol is proven to be correct in isolation, its behavior may 
be unpredictable in an operational network, where interaction with other
protocols and the presence of failures may affect its 
operation.
Protocol errors may be very costly to repair if discovered after 
deployment.
Hence, endeavors should be made to capture protocol 
flaws early in the design cycle before deployment.
To provide an effective solution to the above problems, we present a 
framework for the systematic design and testing of multicast
protocols. 
The framework integrates test generation algorithms with simulation 
and implementation.
We propose a suite of practical methods and tools for automatic
test generation for network protocols.

Many researchers~\cite{formal_survey1}~\cite{formal_survey2}
have developed protocol verification methods to ensure 
certain properties of protocols, like freedom from 
deadlocks or unspecified receptions.
Much of this work, however, was based on 
assumptions about the network conditions, that may not always hold in 
today's Internet, and hence may become invalid.
Other approaches, such as reachability analysis, attempt to check the 
protocol state space, and generally suffer from the `state explosion' 
problem. This problem is exacerbated with the increased complexity of 
the protocol.
Much of the previous work on protocol verification targets
correctness. We target protocol performance and robustness in
the presence of network failures.
In addition, we provide new methods for studying multicast
protocols and topology synthesis that previous works do not
provide.

We investigate two approaches for test generation.
The first approach, called the fault-independent test generation, uses a forward 
search algorithm to explore a subset of the protocol state space to 
generate the test events automatically. 
State and fault equivalence relations are used in this approach to 
reduce the state space.
The second approach is called the fault-oriented test generation, and uses a
mix of forward and backward search techniques to synthesize test events and 
topologies automatically.

We have applied these methods to multicast routing.
Our case studies revealed several design errors, for which we have 
formulated solutions with the aid of this systematic process.

We further suggest an extension of the model
to include end-to-end delays using the notion of virtual LAN. 
Such extension, in conjunction with the
fault-oriented test generation, can be used for performance evaluation of
end-to-end multipoint protocols. 

The rest of this document is organized as follows.
Section~\ref{related} presents 
related work in protocol verification, conformance testing and
VLSI chip testing.
Section~\ref{approach} introduces the proposed framework, and
system definition. 
Sections~\ref{search},~\ref{forward},~\ref{fotg} present the
search based approaches and problem complexity, the
fault-independent test generation and the fault-oriented test
generation, respectively.
Section~\ref{conclusion} concludes~\footnote{We include appendices for
completeness.}.


\begin{itemize}
\item {\bf Multicast Routing Overview}

Multicast protocols are the class of protocols that support group 
communication. 
Multicast routing protocols include, DVMRP~\cite{DVMRP},
MOSPF~\cite{mospf}, PIM-DM~\cite{PIM-DM-SPEC}, CBT~\cite{CBT-sigcomm93}, 
and PIM-SM~\cite{PIM-ARCHv2}.
Multicast routing aims to deliver packets efficiently to group members by 
establishing distribution trees. Figure~\ref{mcast} shows a very simple 
example of a source {\bf $S$} sending to a group of receivers {\bf $R_i$}.

\begin{figure}[th]
 \begin{center}
  \epsfig{file=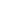,height=8cm,width=6cm,clip=,angle=270}
   \caption{Establishing multicast delivery tree}\label{mcast}
 \end{center}
\end{figure}

Multicast distribution trees may be established by either
broadcast-and-prune or explicit join protocols.
In the former, such as DVMRP or PIM-DM, a multicast packet is broadcast
to all leaf subnetworks. Subnetworks with no local members for the group
send {\em prune} messages towards the source(s) of the packets to stop
further broadcasts.
Link state protocols, such as MOSPF, broadcast membership information
to all nodes.
In contrast, in explicit join protocols, such as CBT or
PIM-SM, routers send hop-by-hop {\em join} messages for the groups and 
sources for which they have local members.

We conduct robustness case studies for PIM-DM.
We are particularly interested in multicast routing protocols,
because they are vulnerable to failure modes, such as selective
loss, that have not been traditionally studied in the area of
protocol design.

For most multicast protocols, when routers are connected via a 
multi-access network (or LAN)\footnote{We use the term LAN to
designate a connected network with respect to IP-multicast. This
includes shared media (such as Ethernet, or FDDI), hubs,
switches, etc.}, hop-by-hop messages are multicast on the 
LAN, and may experience selective loss; i.e. may be received by some 
nodes but not others.
The likelihood of selective loss is increased by the fact that LANs often 
contain hubs, bridges, switches, and other network devices.
Selective loss may affect protocol robustness.

Similarly, end-to-end multicast protocols and applications must deal with 
situations of selective loss. This differentiates these applications 
most clearly from their unicast counterparts, and raises interesting 
robustness questions.

Our case studies illustrate why selective loss should be considered
when evaluating protocol robustness.
This lesson is likely to extend to the design of higher layer protocols 
that operate on top of multicast and can have similar selective loss.

\end{itemize}


\section{Framework Overview}
\label{approach}


Protocols may be evaluated for correctness or performance. 
We refer to
correctness studies that are conducted in
the absence of network failures as verification.
In contrast, robustness studies consider the
presence of network failures (such as packet loss or crashes).
In general,
the robustness of a protocol is its ability to respond correctly
in the face of network component failures and packet loss.  This
work presents a methodology for studying and evaluating
multicast protocols, specifically addressing robustness and
performance issues. 
We propose a framework that integrates automatic test generation
as a basic component for protocol design, along with protocol 
modeling, simulation and implementation testing.
The major contribution of this work lies in developing new
methods for generating stress test scenarios that
target robustness and correctness violation, or worst
case performance.

Instead of studying protocol behavior in isolation, we 
incorporate the protocol model with network dynamics and failures in 
order to reveal more realistic behavior of protocols in operation.

This section presents an overview of the framework and its
constituent components.
The model used to represent the protocol and the system is presented 
along with definitions of the terms used.


Our framework integrates test generation with simulation and 
implementation code. It is used for {\em {\bf S}ystematic {\bf
T}esting of {\bf R}obustness by {\bf E}valuation of {\bf S}ynthesized
{\bf S}cenarios (STRESS)}.
As the name implies, systematic methods for scenario synthesis are a core 
part of the framework. We use the term scenarios to denote the test-suite 
consisting of the topology and events.

The input to this framework is the specification of a protocol, 
and a definition of its design requirements, in terms of
correctness or performance. 
Usually robustness is defined in terms of network dynamics or
fault models.
A fault model represents various component faults; such as packet loss, 
corruption, re-ordering, or machine crashes.
The desired output is a set of test-suites that stress the protocol 
mechanisms according to the robustness criteria.

\begin{figure}[t] 
 \begin{center}
\hspace*{-0.5cm}
  \epsfig{file=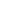,height=10cm,width=7cm,clip=,angle=270}
   \caption{The STRESS framework}\label{framework_fig}
 \end{center}
\end{figure}

As shown in Figure~\ref{framework_fig}, the STRESS framework includes 
test generation, detailed simulation driven by the synthesized tests, and
protocol implementation driven through an emulation interface to the 
simulator. In this work we focus on the test generation (TG) component. 

\subsection{Test Generation}
\label{test_generation}

The core contribution of our work lies in the development of systematic 
test generation algorithms for protocol robustness.
We investigate two such algorithms, each using a different approach. 

In general test generation may be random or deterministic.
Generation of random tests is simple but
a large set of tests is needed to achieve a high measure of
error coverage.
Deterministic test generation (TG), on the other hand, produces tests based on a model
of the protocol. 
The knowledge built into the protocol model enables
the production of shorter and higher-quality test sequences.
Deterministic TG can be: a) fault-independent, or b) fault-oriented.   
Fault-independent TG works without targeting individual faults as 
defined by the fault model. 
Such an approach may employ a forward search technique to inspect the 
protocol state space (or an equivalent subset thereof), after integrating 
the fault into the protocol model.
In this sense, it may be considered a variant of reachability analysis.
We use the notion of equivalence to reduce the search complexity.
Section~\ref{forward} describes our fault-independent approach.

In contrast, fault-oriented tests are generated for specified faults.
Fault-oriented test generation starts from the fault (e.g. a
lost message) and synthesizes the necessary topology and sequence of   
events that trigger the error. This algorithm uses a mix of forward and
backward searches.
We present our fault-oriented algorithm in Section~\ref{fotg}.

We conduct case studies for the multicast routing protocol PIM-DM to illustrate
differences between the approaches, and provide a basis for comparison.

In the remainder of this section, we describe the system
model and definition. 


\subsection{The system model}
\label{system_model}

We define our target system in terms of network and topology elements
and 
a fault model.

\subsubsection{Elements of the network}

Elements of the network consist of multicast capable nodes 
and bi-directional symmetric links. Nodes run same multicast
routing, but not necessarily the same unicast routing.
The topology is an $N$-router LAN modeled at the network level;
we do not model the MAC layer.

For end-to-end performance evaluation, the multicast distribution
tree is abstracted out as delays between end systems and patterns
of loss for the multicast messages.
Cascade of LANs or uniform topologies are addressed in future
research.
                
\subsubsection{The fault model}

We distinguish between the terms {\em error} and {\em fault}.
An {\em error} is a failure of the protocol as defined in the protocol
design requirement and specification. For example, duplication in packet 
delivery is an error for multicast routing.
A {\em fault} is a low level (e.g. physical layer) anomalous behavior,
that may affect the behavior of the protocol under test.
Note that a fault 
may not necessarily be an error for the low level protocol.

The fault model may include: (a) Loss of packets, such as
                packet loss due to congestion or link failures.
		We take into consideration selective packet loss,
where a multicast packet may be received by some members of the group
but not others,
(b) Loss of state, such as multicast and/or unicast
routing tables, due to machine crashes or insufficient memory resources,
(c) The delay model, such as transmission,
                propagation, or queuing delays.
For end-to-end multicast protocols, the delays are those of
the multicast distribution tree and depend upon the multicast routing
protocol, and 
(d) Unicast routing anomalies, such as route 
inconsistencies, oscillations or flapping.

Usually, a fault model is defined in conjunction with the robustness 
criteria for the protocol under study. 
For our robustness studies we study PIM.
The designing robustness goal for PIM is to be able to recover
gracefully (i.e. without going into erroneous stable states) from
single protocol message loss.
That is, being robust to a single message loss 
implies that transitions cause the protocol to move from one
correct stable state to another, even in the presence of selective
message loss. 
In addition, we study PIM protocol behavior in presence of crashes and
route inconsistencies.

\subsection{Test Sequence Definition}

A fault model may include a single fault or multiple faults.
For our robustness studies we adopt a single-fault model,
where only a single fault may occur during a scenario or a test
sequence.

We define two sequences, $T  = <e_1, e_2, \dots, e_n>$ 
and $T' = <e_1, e_2, \dots, e_j, f, e_k, \dots, e_n>$, 
where $e_i$ is an event and $f$ is a fault.
Let $P(q,T)$ be the sequence of states and stimuli of protocol
$P$ under
test $T$ starting from the initial state $q$.
$T'$ is a test sequence if
final $P(q,T')$ is incorrect; i.e. the stable state reached after
the occurrence of the fault does not satisfy the protocol
correctness 
conditions (see Section~\ref{dm_description}) irrespective of
$P(q,T)$.
In case of a fault-free sequence, where $T = T'$, the error is 
attributed to a protocol design error. 
Whereas when $T \neq T'$, and final $P(q,T)$ is correct, 
the error is manifested by the fault.
This definition ignores transient protocol behavior.
We are only concerned with the stable (i.e. non-transient) behavior 
of a protocol.

\subsection{Test Scenario}
\label{test_pattern_definition}

\begin{figure}[th]
 \begin{center}
  \epsfig{file=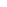,height=7cm,width=4cm,angle=270,clip=}
\vspace*{-.1in}
   \caption{Test pattern dimensions}\label{test_dimensions}
\vspace*{-.1in}
 \end{center}
\end{figure}

A test scenario is defined by a sequence of (host) events,
a topology, and a fault model, as shown in Figure~\ref{test_dimensions}.

The events are actions performed by the host and act as input to the system; for
example, join, leave, or send packet.
The topology is the routed topology of set of nodes and links.
The nodes run the set of protocols under test or other supporting protocols.
The links can be either point-to-point links or LANs.  
This model may be extended later to represent various delays and bandwidths 
between pairs of nodes, by using a virtual LAN matrix (see~\cite{e2e}).
The fault model used to inject the fault into the test. 
According to our single-message loss model, for example, a fault may 
denote the `loss of the second message of type $prune$ traversing a certain link'.
Knowing the location and the triggering action of the fault is important 
in analyzing the protocol behavior.

\subsection{Brief description of PIM-DM}
\label{dm_description}

For our robustness studies, we apply our automatic test generation
algorithms to a version of the Protocol Independent Multicast-Dense
Mode, or PIM-DM.
The description given here is useful for Sections~\ref{search}
through~\ref{fotg}.
   
PIM-DM uses broadcast-and-prune to establish the multicast distribution
trees. In this mode of operation, a multicast packet is broadcast to all
leaf subnetworks. Subnetworks with no local members send {\em prune}
messages towards the source(s) of the packets to stop further broadcasts.

Routers with new members joining the group trigger {\em Graft} messages
towards previously pruned sources to re-establish the branches of the
delivery tree. {\em Graft} messages are acknowledged explicitly at each
hop using the {\em Graft-Ack} message.

PIM-DM uses the underlying unicast routing tables to get the next-hop 
information needed for the RPF (reverse-path-forwarding) checks. This may
lead to situations where there are multiple forwarders for a LAN.
The {\em Assert} mechanism prevents these situations and ensures there is
at most one forwarder for a LAN.

The correct function of a multicast routing protocol in general,
is to deliver data from senders to group members (only those that
have joined the group) without any data loss. For our methods, we
only assume that a correctness definition is given by the protocol
designer or specification. For illustration, we discuss the protocol
errors and the correctness conditions.

\subsubsection{PIM Protocol Errors}

In this study we target protocol design and specification errors.   
We are interested mainly in erroneous stable (i.e. non-transient) states.
In general, the protocol errors may be defined in terms of the end-to-end behavior as
functional correctness requirements.
In our case, for PIM-DM, an error may manifest itself in one of the following ways:

1) {\em black holes}: consecutive packet loss between
periods of packet delivery, 
2) {\em packet looping}: the same packet traverses the same
set of links multiple times,
3) {\em packet duplication}: multiple copies of the same packet are
received by the same receiver(s),
4) {\em join latency}: lack of packet delivery after a receiver
joins the group,
5) {\em leave latency}: unnecessary packet delivery after a receiver 
leaves the group~\footnote{Join and leave latencies may be considered 
in other contexts as performance issues. However, in our study we
treat them as errors.}, and 
6) {\em wasted bandwidth}: unnecessary packet delivery to network
links that do not lead to group members.
   
\subsubsection{Correctness Conditions}
\label{correctness_conditions}

We assume that correctness conditions are provided by the protocol
designer or the protocol specification.
These conditions are necessary to avoid the above protocol errors in a 
LAN environment, and include~\footnote{These are the correctness conditions for
stable states; i.e. not during
transients, and are defined in terms of protocol states (as opposed to
end point behavior). 

The mapping from functional correctness requirements for multicast
routing to the definition in terms of the protocol model is
currently done by the designer. The automation of this process is part
of future research.}:

\begin{enumerate}
\item If one (or more) of the routers is expecting to receive packets
from the LAN, then one other router
must be a forwarder for the LAN.
Violation of this condition may lead to data loss (e.g. join 
latency or black holes).
\item The LAN must have at most one forwarder at a time.
Violation of this condition may lead to data packet duplication.
\item The delivery tree must be loop-free:
\begin{enumerate}
\item Any router should accept packets from one incoming
interface only for each routing entry. This condition is enforced by the RPF
(Reverse Path Forwarding) check.
\item The underlying unicast topology should be loop-free~\footnote{Some
esoteric scenarios of route flapping may lead to multicast loops, in
spite of RPF checks. Currently, our study does not address this issue, as it
does not pertain to a localized behavior.}.
\end{enumerate}
Violation of this condition may lead to data packet looping.  
\item If one of the routers is a forwarder for the LAN, then there must
be at least one router expecting packets from the LANs.
Violation of this condition may lead to leave latency.
\end{enumerate}

\section{Search-based Approaches}
\label{search}

The problem of test synthesis can be viewed as a search problem. By 
searching the possible sequences of events and faults over network
topologies and checking for design requirements (either correctness or 
performance), we can construct the test scenarios that stress the protocol.
However, due to the state space explosion, techniques must be used to 
reduce the complexity of the space to be searched.
We attempt to use these techniques to achieve high
test quality and protocol coverage.


Following we present the GFSM model for the case study 
protocol (PIM-DM), and use it as an illustrative example to analyze the 
complexity of the state space and the search problem, as well as 
illustrate the algorithmic details and principles involved in FITG and FOTG.

\subsection{The Protocol Model}
\label{fsm}

We represent the protocol as a finite state machine (FSM) and the  
overall LAN system by a global FSM (GFSM).

{\em I. FSM model:}
Every instance of the protocol, running on a single router,
is modeled by a deterministic FSM consisting of: (i) a set of
states, (ii) a set of stimuli causing state transitions, and
(iii) a state transition function (or table) describing the state
transition rules. For a system $i$, this is represented by the
machine
${\mathcal{M}}_{i} = ({\mathcal{S}},\tau_{i},\delta_{i})$,
where
$\mathcal{S}$ is a finite set of state symbols,
$\tau_{i}$ is the set of stimuli, and
$\delta_{i}$ is the state transition function
${\mathcal{S}} \times \tau_{i} \rightarrow \mathcal{S}$.

{\em II. Global FSM model:}
The global state is defined as the composition of individual
router states. The output messages from one router may become input 
messages to other routers. Such interaction is captured by the GFSM 
model in the global transition table. The behavior of a 
system with $n$ routers may be described by $\mathcal{M}_{\mathcal{G}} =
(\mathcal{S}_{\mathcal{G}},\tau_{\mathcal{G}},\delta_{\mathcal{G}})$,
where
$\mathcal{S}_{\mathcal{G}}$: ${\mathcal{S}}_{1} \times
{\mathcal{S}}_{2}
\times \dots \times {\mathcal{S}}_{n}$ is the global state space,
$\tau_{\mathcal{G}}$: $\overset{n}{\underset{i=1}{\bigcup}}
\tau_i$ is the set of stimuli, and
$\delta_{\mathcal{G}}$ is the global state transition function
$\mathcal{S}_{\mathcal{G}} \times \tau_{\mathcal{G}}
\rightarrow \mathcal{S}_{\mathcal{G}}$.

The fault model is integrated into the GFSM model. For message loss, the 
transition caused by the message is either nullified or modified, 
depending on the selective loss pattern. Crashes may be treated as 
stimuli causing the routers affected by the crash to transit into a
$crashed$ state~\footnote{The $crashed$ state maybe one of the states 
already defined for the protocol, like the $empty$ state, or may be a 
new state that was not defined previously for the protocol.}. 
Network delays are modeled (when needed) through the delay matrix 
presented in Section~\ref{conclusion}.

\subsection{PIM-DM Model}
\label{fotg_model}

Following is the model of a simplified version of PIM-DM.

\subsubsection{FSM model ${\mathcal{M}}_{i} = 
({\mathcal{S}}_{i},{\tau_i},{\delta_i})$}

For a given group and a given source (i.e., for a specific source-group 
pair), we define the states w.r.t. a specific
LAN to which the router $R_i$ is attached.
For example, a state may indicate that a router is a forwarder
for (or a receiver expecting packets from) the LAN.

\paragraph{System States ($\mathcal{S}$)}

Possible states in which a router may exist are:

\footnotesize

\vspace*{.25cm}
\begin{tabular}{ll} \hline
{\bf State Symbol}   & {\bf Meaning} \\ \hline
$F_i$           & Router $i$ is a forwarder for the LAN \\
$F_{i\_Timer}$   & $i$ forwarder with Timer $_{Timer}$ running \\
$NF_i$          & Upstream router $i$ a non-forwarder \\
$NH_i$          & Router $i$ has the LAN as its next-hop \\
$NH_{i\_Timer}$ & same as $NH_i$ with Timer $_{Timer}$ running \\
$NC_i$          & Router $i$ has a negative-cache entry \\
$EU_i$          & Upstream router $i$ is empty \\
$ED_i$          & Downstream router $i$ is empty \\
$M_i$           & Downstream router with attached member \\
$NM_i$          & Downstream router with no members \\
\hline
\end{tabular}

\normalsize
\renewcommand{\baselinestretch}{1.3}

The possible states for {\em upstream} and {\em downstream}
routers are
as follows:

\begin{equation}
{\mathcal{S}}_{i} =
   \begin{cases}
        \{F_i,F_{i\_Timer},NF_i,EU_i\},& \\ \text{\ \ \ \ if the router is
upstream};
\notag \\
        \{NH_i,NH_{i\_Timer},NC_i,M_i,NM_i,ED_i\},  & \\ \text{\ \ \ \ if the
router is
downstream}.
   \end{cases}
\end{equation}

\renewcommand{\baselinestretch}{1.3}

\paragraph{Stimuli ($\tau$)}

The stimuli considered here include transmitting and
receiving protocol messages, timer events, and external host
events. Only
stimuli leading to change of state
are considered. For example, transmitting messages per se (vs.
receiving
messages) does not cause any
change of state,
except for the $Graft$, in which case the $Rtx$ timer is set.
Following are the stimuli considered in our study:

1. Transmitting messages: Graft transmission ($Graft_{Tx}$).

2. Receiving messages: Graft reception ($Graft_{Rcv}$), Join
reception
($Join$), Prune reception ($Prune$), Graft Acknowledgement
reception
($GAck$), Assert reception ($Assert$), and forwarded packets
reception
($FPkt$).

3. Timer events: these events occur due to timer expiration
($Exp$)
and include the Graft re-transmission timer ($Rtx$), the event of
its
expiration ($RtxExp$), the forwarder-deletion timer ($Del$), and
the event of its expiration ($DelExp$).
We refer to the event of timer expiration as ($Timer Implication$).

4. External host events ($Ext$): include host
sending packets ($SPkt$), host joining a group ($HJoin$ or $HJ$),
and host
leaving a group ($Leave$ or $L$).

$\tau = \{Join,Prune,Graft_{Tx},Graft_{Rcv},GAck,
Assert,$ \\
$FPkt,Rtx,Del,SPkt,HJ,L\}$.

\subsubsection{Global FSM model}

Subscripts are added to distinguish different routers. These subscripts 
are used to describe router semantics and how routers interact on a LAN.
An example global state for a topology of 4 routers connected to
a LAN, with 
router 1 as a forwarder, router 2 expecting packets from the LAN,
and
routers 3 and 4 have negative caches, is given by
$\{F_1, NH_2, NC_3, NC_4\}$.
For the global stimuli $\tau_{\mathcal{G}}$, subscripts are added to stimuli
to denote their originators and recipients (if any). The global transition
rules $\delta_{\mathcal{G}}$ are extended to encompass the router and stimuli
subscripts~\footnote{Semantics of the global stimuli and global transitions 
will be described as needed (see Section~\ref{fotg}).}.

\subsection{Defining stable states}
\label{stable_states}

We are concerned with stable state (i.e. non-transient)
behavior, defined in this section.
        To obtain erroneous stable states, we need to define the
transition mechanisms between such states. We introduce the
concept of transition classification and completion to
distinguish between transient and stable states.

\subsubsection{Classification of Transitions}

We identify two types of transitions; {\em externally
triggered (ET)} and {\em internally triggered (IT)} transitions. 
The
former is stimulated by events external to the system (e.g.,
$HJoin$ or $Leave$), whereas the latter is stimulated by
events internal to the system (e.g., $FPkt$ or $Graft$).

We note that some transitions may be triggered due to either
internal and
external events, depending on the scenario.
For example, a $Prune$ may be triggered due to forwarding packets
by an
upstream router $FPkt$ (which is an internal event), or a $Leave$
(which   
is an external event).

A global state is checked for correctness at the end of
an externally triggered transition after completing its dependent
internally triggered transitions.

Following is a table of host events, their dependent ET 
and IT events:

\footnotesize
   
\begin{tabular}{|l|l|l|l|} \hline
{\bf Host Events}& $SPkt$ & $HJoin$ & $Leave$ \\ \hline
{\bf ET events}       & $FPkt$ & $Graft$ & $Prune$ \\ \hline
{\bf IT events}       & $Assert$, $Prune$, & $GAck$ & $Join$ \\
                & $Join$          &       & \\ \hline
\end{tabular}

\normalsize
\renewcommand{\baselinestretch}{1.3}
        
\subsubsection{Transition Completion}
\label{transition_completion}

To check for the global system correctness, all stimulated
internal transitions should be completed, to bring the system 
into a stable state.
Intermediate (transient) states should not be checked
for correctness (since they may temporarily seem to violate 
the correctness conditions set forth for stable states, and 
hence may give false error indication).
The process of identifying complete transitions depends on the
nature of the protocol.
But, in general, we may identify a complete transition sequence, 
as the
sequence of (all) transitions triggered due to a single external 
stimulus 
(e.g., $HJoin$ or $Leave$).
Therefore, we should be able to identify a transition based upon
its stimuli (either external or internal).
At the end of each complete transition sequence the system exists
in either a correct or erroneous stable state.
Event-triggered timers (e.g., $Del$, $Rtx$) fire at the end of a
complete transition.

\subsection{Problem Complexity}
\label{complexity}

The problem of finding test scenarios leading to protocol 
error can be viewed as a search problem of
the protocol state space. Conventional
reachability analysis~\cite{reachability} attempts
to investigate this space exhaustively
and incurs the 'state space explosion'
problem. To circumvent this problem we use search
reduction techniques using domain-specific information
of multicast routing.

In this section, we give the complexity of
exhaustive search, then discuss the reduction techniques we employ
based on notion of equivalence, and give the complexity of the
state space.

\subsubsection{Complexity of exhaustive search}   

Exhaustive search attempts to generate
all states reachable from initial system states.
For a system of $n$ routers where each router may exist in any
state $s_i \in S$, and $|{\mathcal{S}}|= s$ states,
the number of reachable states in the system is bounded by
$(s)^n$. 
With $l$ possible transitions we need $l \cdot (s)^n$ state visits to
investigate all transitions.
Faults, such as message loss and crashes, increase the 
branching factor $l$, and may introduce new states increasing 
${\mathcal{S}}$. 
For our case study $|{\mathcal{S}}|= 10$, while selective loss and
crashes~\footnote{Crashes force any state to the empty state.}
increase branching almost by factor of 9.

\subsubsection{State reduction through equivalence}
\label{reduction}

Exhaustive search has exponential complexity.
To reduce this complexity we use the notion of equivalence.
Intuitively, in multicast routing the order in which the states are considered
is irrelevant (e.g., if router $R_1$ or $R_4$ is a
forwarder is insignificant, so long as there is only one forwarder).
Hence, we can treat the global state as an unordered set of state symbols. 
This concept is called `counting equivalence'~\footnote{Two system states
($q_1,q_2,\dots ,q_n$) and ($p_1,p_2,\dots ,p_n)$ are strictly
equivalent iff
$q_i = p_i$, where $q_i ,p_i \in {\mathcal{S}}, \forall 1 \leq i \leq n$.
However, all routers use the same deterministic FSM model, 
hence all $n!$ permutations of ($q_1,q_2,\dots ,q_n$) are  
equivalent.
A global state for a system with $n$ routers may be represented as
$\sideset{}{_{i=1}^{|\mathcal{S}|}}\prod s_i^{k_i}$, where $k_i$
is the number of routers in state $s_i \in \mathcal{S}$ and
$\sideset{}{_{i=1}^{|\mathcal{S}|}}\Sigma k_i = n$.
Formally, {\em Counting Equivalence} states that
{\em two system states
$\sideset{}{_{i=1}^{|\mathcal{S}|}}\prod s_i^{k_i}$ and
$\sideset{}{_{i=1}^{|\mathcal{S}|}}\prod s_i^{l_i}$ are
equivalent
if $k_i = l_i \forall i$}.}.
By definition, the notion of equivalence implies that by
investigating the equivalent subspace we can test for protocol
correctness. That is, if the equivalent subspace is verified to
be correct then the protocol is correct, and if there is an error
in the protocol then it must exist in the equivalent subspace~\footnote{The notion of
counting equivalence also applies to transitions and faults. 
Those transitions or faults leading to equivalent states are considered 
equivalent.}.

\paragraph{\bf Symbolic representation}
\label{symbolic}

We use a symbolic representation as a convenient form of representing the
global state to illustrate the notion of equivalence and to help in
defining the error and correct states in a succinct manner.
In the symbolic representation, $r$ routers in state $q$ are
represented by
$q^r$.
The global state for a system of $n$ routers is represented by
$G = \left( q_1^{r_1},q_2^{r_2},\dots ,q_m^{r_m} \right)$, where
$m = |\mathcal{S}|$, $\Sigma r_i = n$. 
For symbolic representation of topologies where $n$ is unknown $r_i \in
[0,1,2,1+,*]$ (`1+' is 1 or more, and `*' is 0 or more).

To satisfy the correctness conditions for PIM-DM,
the correct stable global  
states are
those containing no forwarders and no routers expecting packets,
or those
containing one forwarder and one or more routers expecting
packets from
the link; symbolically this may be given by: 
$G_1 = \left( F^0, NH^0, NC^* \right)$, and
$G_2 = \left( F^1, NH^{1+}, NC^* \right)$.~\footnote{For
convenience, we may represent these two states as
$G_1 = \left( NC^* \right)$, and
$G_2 = \left( F, NH^{1+}, NC^* \right)$.}

We use $X$ to denote $any$ state $s_i \in {\mathcal{S}}$. For
example, $\{X - F\}^*$ denotes 0 or more states
$s_i \in {\mathcal{S}} - \{F\}$.
This symbolic representation is used
to estimate the size of the reduced state space.

\paragraph{\bf Complexity of the state space with equivalence
reduction}

Considering counting equivalence, finding the number of
equivalent states becomes a problem of combinatorics.
The number of equivalent states becomes
$C(n+s-1,n) = \frac{(n+s-1)!}{n! \cdot (s-1)!}$ ,
where, $n$ is the number of routers, $s$ is the number of state symbols,
and $C(x,y) = \frac{x!}{y! \cdot (x-y)!}$, is the number of   
$y$-combination of $x$-set~\cite{algorithms_book}.

\subsubsection{Representation of error and correct states}
\label{correctness}

Depending on the correctness definition we may get different
counts for the number of correct or error states.
To get an idea about the size of the correct or error state space
for our case study, we take two definitions of correctness and  
compute the number of correct states.
For the correct states of PIM-DM,
we either have: (1) no forwarders with no routers expecting
packets from the LAN, or (2) exactly one
forwarder with routers expecting packets from the LAN~\footnote{These conditions we
have found to be reasonably sufficient to meet the 
functional correctness requirements. However, they may not be necessary, 
hence the search may generate false errors. Proving necessity is part of 
future work.}.

The correct space and the erroneous space must be disjoint and
they must be complete (i.e. add up to the complete space),
otherwise the specification is incorrect. See Appendix I-A 
for details.

We present two correctness definitions that are
used in our case.

\begin{itemize}

\item The first definition considers the forwarder 
states as $F$ and the routers
expecting packets from the LAN as $NH$. Hence, the
symbolic representation of the correct states becomes:
$\left( \{X-NH-F\}^* \right)$, or
$\left( NH, F, \{X-F\}^* \right)$,

and the number of correct states is: $C(n+s-3,n) + C(n+s-4,n-2).$

\item The second definition considers the forwarder states
as $\{F_i, F_{i\_{Del}}\}$ or simply $F_X$, and the states
expecting packets from the LAN as $\{NH_i, NH_{i\_{Rtx}}\}$ or
simply $NH_X$.
Hence, the symbolic representation of the correct states becomes:
$\left( \{X-NH_X-F_X\}^* \right)$, or
$\left( NH_X, F_X, \{X-F_X\}^* \right)$,

and the number of correct states is:

$C(n+s-5,n) + 4 \cdot C(n+s-5,n-2) - 2 \cdot C(n+s-6,n-3).$

\end{itemize}

Refer to Appendix I-B 
for more details on
deriving the number of correct states.

In general, we find that the size of the error state space,
according to both definitions, constitutes the major portion of
the whole state space. This means that search techniques
explicitly exploring the
error states are likely to be more complex than others. We take
this in consideration when designing our methods.

\section{Fault-independent Test Generation}
\label{forward}

Fault-independent test generation (FITG) uses the forward search 
technique to investigate parts of the state space.
As in reachability analysis, forward search starts from initial
states and applies the stimuli repeatedly to produce the
reachable state space (or part thereof). Conventionally, an exhaustive 
search is conducted to explore the state space.
In the exhaustive approach all reachable states are expanded until the 
reachable state space is exhausted. We use several manifestations of the 
notion of counting equivalence introduced earlier to reduce the 
complexity of the exhaustive algorithm and expand only equivalent subspaces.
To examine robustness of the protocol, we incorporate selective loss 
scenarios into the search.	

\subsection{Reduction Using Equivalences}

The search procedure starts from the initial states~\footnote{For our case study the
routers start as either a non-member ($NM$) or empty upstream routers ($EU$), that is,
the initial states $I.S. = \{NM,EU\}$.}
and keeps a list of states visited to prevent looping.
Each state is expanded by applying the stimuli and advancing the state 
machine forward by implementing the transition rules and returning 
a new stable state each time~\footnote{For details of the above procedures, see
Appendix II-A.}. 
We use the counting equivalence notion
to reduce the complexity of the search in three stages of the
search:
\begin{enumerate}
\item The first reduction we use is to investigate only the
equivalent initial states. To achieve this we simply treat the set of 
states constituting the global state as unordered set instead of ordered 
set.
For example, the output of such procedure for $I.S. = \{NM,EU\}$ and $n=2$ 
would be: $\{NM,NM\},\{NM,EU\},\{EU,EU\}$.

One procedure that produces such equivalent initial state space
given in Appendix II-B. 
The complexity of the this algorithm is given by $C(n + i.s. -
1,n)$ as was shown in Section~\ref{reduction} and verified 
through simulation.

\item The second reduction we use is during comparison of visited states.
Instead of comparing the actual states, we compare and store equivalent
states. Hence, for example, the states $\{NF_1, NH_2\}$ and 
$\{NH_1, NF_2\}$ are equivalent.

\item A third reduction is made based on the observation that
applying identical stimuli to different routers in identical
states leads to equivalent global states. 
Hence, we can eliminate some redundant transitions.
For example, for the global state $\{ NH_1, NH_2, F_3 \}$ a
$Leave$ applied to $R_1$ or $R_2$ would
produce the equivalent state $\{NH^1, NC^1, F^1 \}$.
To achieve this reduction we add flag check before advancing the state 
machine forward.
We call the algorithm after the third reduction the {\bf reduced} algorithm.

\end{enumerate} 

In all the above algorithms, a forward step advances the
GFSM to the next stable state. This is done by applying all the
internally dependent stimuli (elicited due to the applied
external stimulus) in addition to any timer implications, if any
exists. Only stable states are checked for correctness.

\subsection{Applying the Method}
\label{fitg_apply}

In this section we discuss how the fault-independent test
generation can be applied to the model of PIM-DM.
We apply forward search techniques to study correctness of PIM-DM. 
We first study the complexity of the algorithms without faults. Then we
apply selective message loss to study the protocol behavior and analyze the 
protocol errors.

\subsubsection{Method input}
\label{fitg_input}

The protocol model is provided by the designer or protocol specification,
in terms of a transition table or transition rules of the GFSM, and a set 
of initial state symbols. The design requirements, in terms of 
correctness in this case, is assumed to be also given by the protocol 
specification. This includes definition of correct states or erroneous 
states, in addition to the fault model if studying robustness. 
Furthermore, the detection of equivalence classes needs to be provided 
by the designer~\footnote{For our case study, the symmetry inherent in
multicast over LANs was used to establish the counting equivalence for
states, transitions and faults.}. 
Currently, we do not automate the detection of equivalent classes. 
Also, the number of routers in the topology or topologies to be 
investigated (i.e., on the LAN) has to be specified.

\subsubsection{Complexity of forward search for PIM-DM}
\label{fitg_complexity}

\begin{figure}[t]  
 \begin{center}    
  \epsfig{file=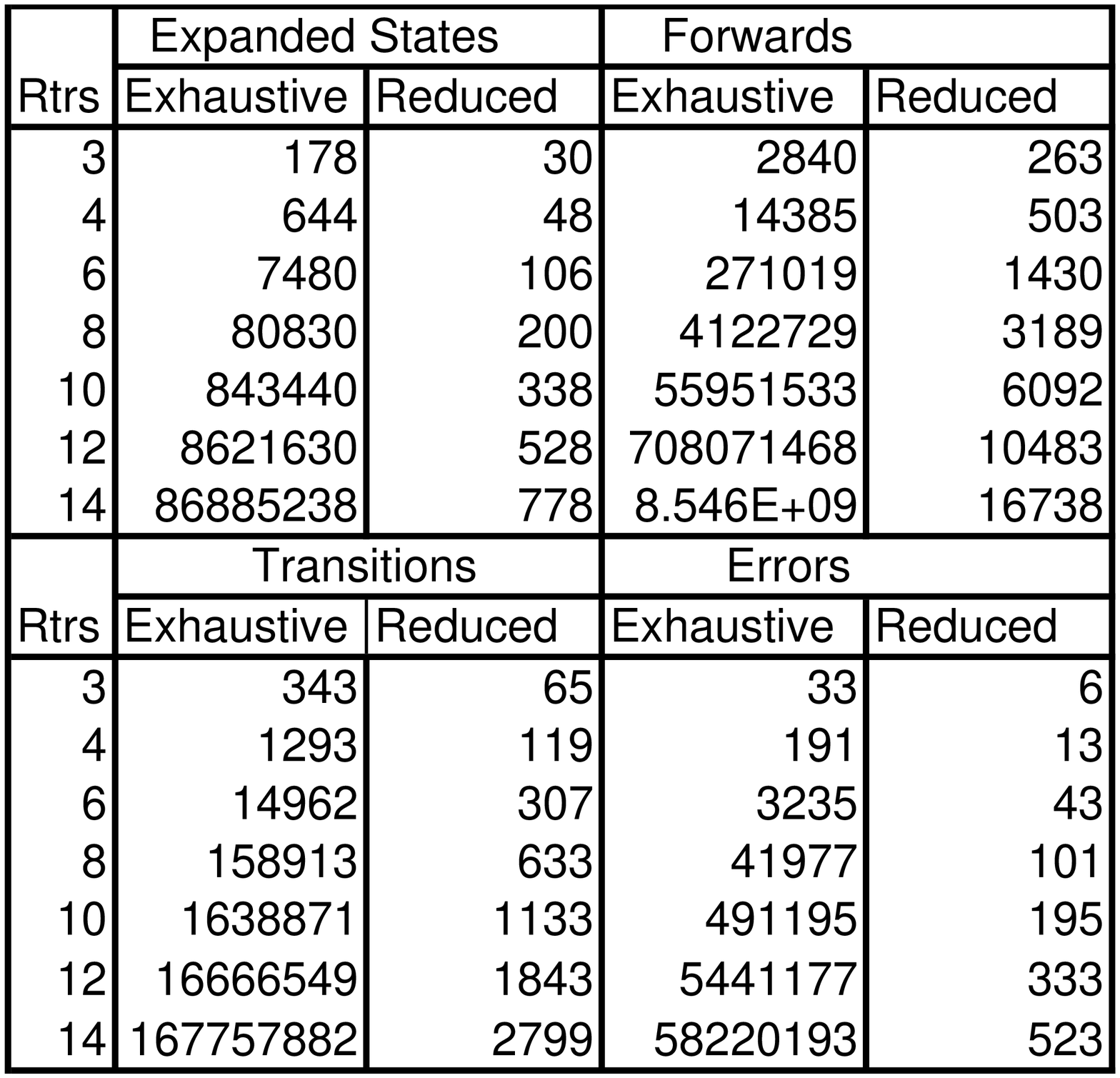,height=4cm,width=8cm,clip=,angle=0}
   \caption{Simulation statistics for
forward algorithms. {\em Expanded States} is the number
of stable states visited, $Forwards$ is the number of
forward advances of the state machine,
$Transitions$ is the number of transient states 
visited and $Errors$ is the number of stable state errors 
detected.}\label{tbl1.1}
 \end{center}   
\end{figure}

The procedures presented above were simulated for PIM-DM
to study its correctness. 
This set of results shows behavior of the algorithms 
without including faults, i.e., when used for verification. 
We identified the initial 
state symbols to be $\{NM,EU\}$; $NM$ for downstream routers and 
$EU$ for upstream routers. The number of reachable states visited, the 
number of transitions and the number of erroneous states found were
recorded. Summary of the results is given in  
Figure~\ref{tbl1.1}.
The number of expanded states denotes the number of visited   
stable states.
The number of `forwards' is
the number of times the state machine was advanced forward 
denoting the number of transitions between stable states. The number of
transitions is the number of visited transient states,
and the number of error states is the number of stable (or
expanded) states violating the correctness conditions. The error
condition is given as in the second error condition in
Section~\ref{correctness}. Note that each of the other error states is
equivalent to at least one error state detected by the {\bf reduced} 
algorithm. Hence,
having less number of discovered error states by
an algorithm in this case does not mean losing any information or
causes of error, which follows from the definition of
equivalence. 
Reducing the error states means reducing the time needed to analyze the 
errors.
 
We notice that there significant reduction in the algorithm
complexity with the use of equivalence relations. In particular,
the number of transitions is reduced from $O(4^n)$ for the
exhaustive algorithm, to $O(n^4)$ for the 
{\bf reduced} algorithm.
Similar results were obtained for the number 
of forwards, expanded states and number of error states.
The reduction gained by using the counting equivalence is exponential.
More detailed presentation of the algorithmic details and results are 
given in Appendix II. 

For robustness analysis (vs. verification), faults are included in 
the GFSM model. Intuitively, an increase in the overall complexity of the 
algorithms will be observed. Although we have only applied faults to 
study the behavior of the protocol and not the complexity of the search, we 
anticipate similar asymptotic reduction gains using counting 
equivalence.

\subsubsection{Summary of behavioral errors for PIM-DM}

We used the above algorithm to search the protocol model for PIM-DM.
Correctness was checked automatically by the method by checking the stable
states (i.e., after applying complete transitions). By analyzing the 
sequence of events leading to error we were able to reason about the protocol
behavior.
Several PIM-DM errors were detected by the method, some 
pertaining to correctness in the absence of message loss, while others 
were only detected in the presence of message loss.
We have studied cases of up to 14-router LANs. Sometimes errors were
found to occur in different topologies for similar reasons as will be
shown.
Here, we only discuss results for the two router and 3-router LAN cases for
illustration.

\begin{itemize}
\item Only one error was detected in the two-router case. With the 
initial state $\{EU,EU\}$ (i.e., both routers are upstream routers), the 
system enters the error state $\{F, NF\}$, where there is a forwarder for 
the LAN but there are no routers expecting packets or attached members. 
In this case the $Assert$ process chose one forwarder for the LAN, but 
there were no downstream routers to $Prune$ off the extra traffic, and so 
the protocol causes wasted bandwidth.
\item Several errors were detected for the 3-router LAN case:
\begin{itemize}
\item Starting from $\{EU,EU,EU\}$ the system enters the error state 
$\{F,NF,NF\}$ for a similar reason to that given above.
\item Starting from $\{NM,EU,EU\}$ the system enters the error state 
$\{NC,NF,F\}$. By analyzing the trace of events leading to the error we 
notice that the downstream router $NC$ pruned off one of the upstream 
routers, $NF$, before the $Assert$ process takes place to choose a winner 
for the LAN. Hence the protocol causes wasted bandwidth.
\item Starting from $\{NM,EU,EU\}$ the system enters state $\{NH,F,F\}$. 
This is due to the transition table rules, when a forwarder sends a 
packet, all upstream routers in the $EU$ state transit into $F$ state. 
This is not an actual error, however, since the system will recover with 
the next forwarded packet using $Assert$~\footnote{This is one case
where the correctness conditions for the model are sufficient but not
necessary to meet the functional requirements for correctness, thus
leading to a false error. Sufficiency and necessity proofs are subject of
future work.}. The detection of this false-error 
could have been avoided by issuing $SPkt$ stimulus before the error 
check, to see if the system will recover with the next packet sent.
\item With message loss, errors were detected for $Join$ and $Prune$ loss.
When the system is in $\{NH,NH,F\}$ state and one of the downstream 
members leaves (i.e., issues $L$ event), a $Prune$ is sent on the LAN. If 
this $Prune$ is selectively lost by the other downstream router, a $Join$ 
will not be sent and the system enters state $\{NC,NH,NF\}$. Similarly, if 
the $Join$ is lost, the protocol ends up in an error state.
\end{itemize}
\end{itemize}

\subsection{Challenges and Limitations}

In order to generalize the fault-independent test generation method, we 
need to address several open research issues and challenges. 

\begin{itemize}

\item The topology is an input to the method in terms of number of routers.
To add topology synthesis to FITG we may use the symbolic representation 
presented in Section~\ref{complexity}, where the use of repetition 
constructs~\footnote{Repetition constructs 
include, for example, the `*' to represent zero or more states, or the 
`1+' to represent one or more states, `2+' two or more, so on.} may be 
used to represent the LAN topology in general.
A similar principle was used in~\cite{cache_coherence} for cache coherence 
protocol verification, where the state space is split 
using repetition constructs based on the correctness definition.
In Section~\ref{fotg} we present a new method that synthesizes 
the topology automatically as part of the search process.

\item Equivalence classes are given as input to the method. In this 
study we have used symmetries inherent in multicast routing on LANs to 
utilize equivalence. This symmetry may not exist in other protocols or 
topologies, hence the forward search may become increasingly complex.
Automating identification of equivalence classes is part of future work.

Other kinds of equivalence may be investigated to reduce complexity in 
these cases~\footnote{An example of another kind of equivalence is {\em fault 
dominance}, where a system is proven to necessarily reach one error before
reaching another, thus the former error dominates the latter error.}. Also,
other techniques for complexity reduction
may be investigated, such as statistical sampling based on randomization or 
hashing used in SPIN~\cite{spin}. However, sampling techniques do not achieve full 
coverage of the state space.

\item The topology used in this study is limited to a single-hop LAN.
Although we 
found it quite useful to study multicast routing over LANs, the method 
needs to be extended to multi-hop LAN to be more general. 
Our work in~\cite{e2e} introduces the notion of $virtual$ LAN, and future 
work addresses multi-LAN topologies.

\end{itemize}

In sum, the fault-independent test generation may be used for protocol 
verification given the symmetry inherent in the system 
studied (i.e., protocol and topology). For robustness studies, where 
the fault model is included in the search, the complexity of the search 
grows. In this approach we did not address performance issues or topology
synthesis. These issues are addressed in the coming sections.
However, we shall re-use the notion of forward search and the use of
counting equivalence in the method discussed next.

\section{Fault-oriented Test Generation}
\label{fotg}


In this section, we investigate the fault-oriented test generation (FOTG), 
where the tests are generated for specific faults. 
In this method, the test generation algorithm starts from the fault(s) and 
searches for a possible error, establishing the necessary topology and 
events to produce the error. 
Once the error is established, a backward search technique produces a 
test sequence leading to the erroneous state, if such a state is reachable.
We use the FSM formalism presented in Section~\ref{search}
to represent the protocol.
We also re-use some ideas from the FITG algorithm previously
presented, such as forward search and the notion of equivalence for
search reduction.

\subsection{FOTG Method Overview}
\label{fotg_overview}

Fault-oriented test generation (FOTG) targets specific faults or 
conditions, and so is better suited to study robustness in the 
presence of faults in general.
FOTG has three main stages: a) topology synthesis, b) forward 
implication and error detection, and c) backward implication.
The topology synthesis establishes the necessary components (e.g., 
routers and hosts) of the system to trigger the given condition (e.g., 
trigger a protocol message). This leads to the formation of a global 
state in the middle of the state space~\footnote{The global state from 
which FOTG starts is synthesized for a given fault, such as a message 
to be lost.}.
Forward search is then performed from that global state in its vicinity, i.e.,
within a complete transition, after applying the fault. 
This process is called {\em forward implication}, and uses search techniques 
similar to those explained earlier in Section~\ref{forward}.
If an error occurs, backward search is performed thereafter to establish a valid
sequence leading from an initial state to the synthesized global
state. To achieve this, the transition rules are reversed and a
search is performed until an initial state is reached, or the
synthesized state is declared unreachable.
This process is called {\em backward implication}.

Much of the algorithmic details are based on $condition \rightarrow
effect$ reasoning of the transition rules. 
This reasoning is emphasized in the semantics of the transition table used
in the topology synthesis and the backward search. 
Section~\ref{transition_table} describes these semantics.
In Section~\ref{fotg_details} we describe the algorithmic details of FOTG,
and in Section~\ref{fotg_apply} we describe how FOTG was applies to PIM-DM
in our case study, and present the results and method evaluation.
Section~\ref{fotg_discuss} we discuss the limitations of the method and our
findings.

\subsubsection{The Transition Table}
\label{transition_table}

The global state transition may be represented in several ways.
Here, we 
choose a transition table representation that emphasizes the
effect of 
the stimuli on the system, and hence facilitates topology
synthesis.
The transition table describes, for each stimulus, the 
conditions of its occurrence.
A condition is given as stimulus and state or transition
(denoted by {\em 
stimulus.state/trans}), where the transition is given as 
$startState \rightarrow endState$.

We further extend message and router semantics to capture multicast 
semantics. Following, we present a detailed description of the
semantics of the transition table then give the resulting transition
table for our case study, to be used later in this section.

\paragraph{Semantics of the transition table}

In this subsection we describe the message and router 
semantics, pre-conditions, and post-conditions.

\begin{itemize}

\item{Stimuli and router semantics:}
Stimuli are classified based on the routers affected by them.
Stimuli types include:

\begin{enumerate} 
\item $orig$: stimuli or events occurring within the router originating the
stimulus but do not affect other routers,
and include $HJ$, $L$,
$SPkt$, $Graft_{Tx}$, $Del$ and $Rtx$.

\item $dst$: messages that are processed by the destination
router only, and include $Join$, $GAck$ and $Graft_{Rcv}$.

\item $mcast$: multicast messages that are processed by all other
routers, and include $Assert$ and $FPkt$.

\item $mcastDownstream$: multicast messages that are processed by
all other downstream routers, but only one upstream router, and
includes the $Prune$ message.
\end{enumerate}

These types are used by the search algorithm for processing the
stimuli and messages.
According to these different types of stimuli processing a
router may take as subscript `$orig$', `$dst$', or `$other$'.
The `$orig$' symbol designates the originating router of the stimulus or message,
whereas `$dst$' designates the destination of the message. 
`$other$' indicates routers other than the originator.
Routers are also classified as $upstream$ or $downstream$ as presented in 
Section~\ref{search}.

\item{Pre-Conditions:}
The pre-conditions in general are of the form
$stimulus.state/transition$,
where the transition is given as $startState \rightarrow endState$.
If there are several pre-conditions, then we can use a logical OR
to represent the rule.
At least one pre-condition is necessary to trigger the stimulus.
Example of a $stimulus.state$ condition is the condition for $Join$ message,
namely,
$Prune_{other}.NH_{orig}$, that 
is, a $Join$ is triggered by the reception of a $Prune$ from another router,
with the originator of the $Join$ in $NH$.
An example of a $stimulus.transition$ condition is the condition for Graft
transmission $HJ.(NC \rightarrow NH)$; i.e. a host joining and
the transition of the router from the negative cache state to the
next hop state.

\item{Post-Conditions:}
A post-condition is an event and/or transition that is triggered
by the stimulus.~\footnote{Network faults, such as message loss,
may cause the stimulus not to take effect. For example, losing a
$Join$ message will cause the event of $Join$ reception not to
take effect.}
Post-conditions may be in the form of: (1) $transition$, 
(2) $condition.transition$, (3) $condition.stimulus$, and 
(4) $stimulus.transition$.

\begin{enumerate}
\item $transition$: has an implicit condition with
which it is associated; i.e. `$a \rightarrow b$' means 
`if $a \in GState$ then $a \rightarrow b$'.
For example, $Join$ post-condition ($NF_{dst} \rightarrow
F_{dst}$), means if $NF_{dst} \in GState$ then
transition $NF \rightarrow F$ will occur.

\item $Condition.transition$: is same as (1) except the condition is
explicit~\footnote{This does not appear in our case study.}.

\item $Condition.stimulus$: if the condition is satisfied then
the stimulus is triggered. 
For example, $Prune$ post-condition
`$NH_{other}.Join_{other}$', means that for all $NH_x \in GState$
(where $x$ is not equal to $orig$) then have router $x$ trigger a
$Join$.

\item $Stimulus.transition$: has the transition condition implied as
in (1) above. For example, $Graft_{Rcv}$ post-condition
`$GAck.(NF_{dst} \rightarrow F_{dst})$', means if $NF_{dst} \in GState$, 
then the transition occurs and $GAck$ is triggered.

\end{enumerate}

If more than one post-condition exists, then the logical
relation between them is either an `XOR' if the router is the 
same, or an `AND' if the routers are different.
For example, $Join$ post-conditions are `$F_{dst\_Del}
\rightarrow F_{dst}, NF_{dst} \rightarrow F_{dst}$',
which means ($F_{dst\_{Del}} \rightarrow F_{dst}$) XOR ($NF_{dst}
\rightarrow F_{dst}$).~\footnote{There is an implicit condition
that can never be satisfied in both statements, which is the
existence of $dst$ in only one state at a time.}

On the other hand, $Prune$ post-conditions are `$F_{dst} \rightarrow 
F_{dst\_{Del}}, NH_{other}.Join_{other}$',
which implies that the transition will occur if $F_{dst} \in
GState$ AND a $Join$ will be triggered if $NH \in GState$.




\end{itemize}

Following is the transition table used in our case study.

\scriptsize

\vspace*{.03in}
\hspace{-.5cm}
\begin{tabular}{|l|l|l|} \hline
{\bf Stimulus} & {\bf Pre-conditions} & {\bf 
Post-conditions} \\ \hline
$Join$ & $Prune_{other}. NH_{orig}$ & $F_{dst\_Del} \rightarrow
F_{dst}, 
NF_{dst}\rightarrow F_{dst}$ \\ \hline
$Prune$ & $L.NC, FPkt.NC$ & 
$F_{dst} \rightarrow F_{dst\_Del},$ \\
& & $NH_{other}.Join_{other}$ \\ 
\hline 
$Graft_{Tx}$ & $HJ.(NC \rightarrow NH),$ 
 & $Graft_{Rcv}.(NH \rightarrow NH_{\_Rtx})$ \\ 
& $Rtx Exp.(NH_{\_Rtx} \rightarrow NH)$ & \\ \hline
$Graft_{Rcv}$ & $Graft_{Tx}.(NH \rightarrow NH_{\_Rtx})$ &
$GAck.(NF_{dst} 
\rightarrow F_{dst})$ \\ \hline
$GAck$ & $Graft_{Rcv}.F$ & $NH_{dst\_Rtx} \rightarrow NH_{dst}$
\\ \hline 
$Assert$ & $FPkt_{other}.F_{orig} $ & $F_{other} 
\rightarrow NF_{other} $ \\ \hline
$FPkt$ & $Spkt.F$ & $Prune.(NM \rightarrow NC),$ \\
& & $ ED \rightarrow NH,M \rightarrow NH,$ \\
& & $EU_{other} \rightarrow F_{other}$, $F_{other}.Assert$ \\ \hline
$Rtx$ & RtxExp & $Graft_{Tx}.(NH_{orig\_Rtx} \rightarrow
NH_{orig})$  
\\ \hline
$Del$ & DelExp & $F_{orig\_Del} \rightarrow NF_{orig}$ \\ \hline
$SPkt$ & Ext & $FPkt.(EU_{orig} \rightarrow F_{orig})$ \\ \hline
$HJoin$ & Ext & $NM \rightarrow M, Graft_{Tx}.(NC \rightarrow
NH)$ \\ \hline
$Leave$ & Ext & $M \rightarrow NM, Prune.(NH \rightarrow NC),$ \\ 
& & $Prune.(NH_{Rtx} \rightarrow NC)$ \\ \hline
\end{tabular}

\normalsize

The above pre-conditions can be derived automatically from the
post-conditions.
In  Appendix III, 
we describe the `PreConditions' procedure that takes as
input one form of the
conventional post-condition transition table and produces the pre-condition
semantics.

\paragraph{State Dependency Table}
\label{dep_table}

To aid in test sequence synthesis
through the backward implication 
procedure, we construct what we call a state dependency 
table. This table can be inferred automatically from 
the transition table.
We use this table to improve the performance of the
algorithm and for illustration.

For each state, the dependency table contains the
possible preceding 
states and the stimulus from which the state can be
reached or implied.
To obtain this information for a state $s$, the algorithm
the post-conditions of the transition 
table for entries where the $endState$ of a transition
is $s$.
In addition, a state may be identified as an initial
state (I.S.), and hence can be readily established without any 
preceding states.
The `dependencyTable' procedure in Appendix III 
generates 
the dependency table from the transition table of conditions.
For $s \in I.S.$ a symbol denoting initial state is added to the
array entry. For our case study $I.S. = \{NM,EU\}$.
Based on the above transition table, following is the resulting 
state dependency table:~\footnote{The possible backward implications are
separated by `commas' indicating `OR' relation.}

\vspace{.2cm}
\hspace{-.4cm}
\footnotesize
\tablehead{\hline}
\tabletail{\hline}
\begin{supertabular}{|l|l|}
{\bf State}   & {\bf Possible Backward Implication(s)} \\ \hline \hline
$F_i$           & $\stackrel{FPkt_{other}}{\longleftarrow}EU_i, 
\stackrel{Join}{\longleftarrow} F_{i\_Del}, 
\stackrel{Join}{\longleftarrow} NF_i, 
\stackrel{Graft_{Rcv}}{\longleftarrow} NF_i,$ \\
& $ \stackrel{SPkt}{\longleftarrow} EU_i$ \\ \hline 
$F_{i\_Del}$    & 
$\stackrel{Prune}{\longleftarrow} F_i$ \\ \hline
$NF_i$          & $\stackrel{Del}{\longleftarrow} F_{i\_Del}, 
\stackrel{Assert}{\longleftarrow}
F_i$ \\ \hline $NH_i$          & 
$\stackrel{Rtx, GAck}{\longleftarrow}
NH_{i\_Rtx}, 
\stackrel{HJ}{\longleftarrow}
NC_i, \stackrel{FPkt}{\longleftarrow} M_i, 
\stackrel{FPkt}{\longleftarrow} ED_i$ \\ \hline
$NH_{i\_Rtx}$	& 
$\stackrel{Graft_{Tx}}{\longleftarrow} NH_i$ \\ \hline
$NC_i$          & $\stackrel{FPkt}{\longleftarrow} NM_i, 
\stackrel{L}{\longleftarrow} NH_{i\_Rtx}, 
\stackrel{L}{\longleftarrow} NH_i$ \\ 
\hline 
$EU_i$          & $\leftarrow I.S.$ \\ \hline 
$ED_i$          & $\leftarrow I.S.$ \\ \hline
$M_i$           & $\stackrel{HJ}{\longleftarrow} NM_i$ \\ \hline
$NM_i$          & $\stackrel{L}{\longleftarrow} M_i, \leftarrow I.S.$ \\
\hline
\end{supertabular}

\normalsize
\vspace{.2cm}

In cases where the stimulus affects more than one router (e.g., multicast
$Prune$), multiple states need to be simultaneously implied
in 
one backward step, otherwise an $I.S.$ may not be reached. To do
this, 
the transitions in the post-conditions of the stimulus are
traversed, 
and any states in the global state that are $endState$s are
replaced 
by their corresponding $startState$s. For example,
$\{M_i,NM_j,F_k\} \stackrel{FPkt}{\longleftarrow} \{NH_i,NC_j,F_k\}$. 
This is taken care of by the backward implication section described later.

\subsection{FOTG details}
\label{fotg_details}

As previously mentioned, our FOTG approach consists of three
phases: 
I) synthesis of the global state to inspect, 
II) forward implication, and 
III) backward implication. 
These phases are explained in more detail in this section.
In Section~\ref{fotg_apply} we present an illustrative example for
the these phases.

\subsubsection{Synthesizing the Global State}
~\\
Starting from a condition (e.g., protocol message or stimulus),
and using  the information in the protocol model (i.e. the
transition table),
a global state is synthesized for investigation. We refer to this
state as the global-state inspected ($G_I$), and it is obtained
as follows:

\begin{enumerate}
\item 
The global state is initially empty and the inspected 
stimulus is initially set to the stimulus investigated.

\item 
 For the inspected stimulus, the state(s) (or the $startState$(s) of
the transition) of the post-condition are obtained from the transition
table.
If these states do not exist in the global state, and cannot be
inferred therefrom, then they are added to the global state.

\item 
 For the inspected stimulus, the state(s) (or the $endState$(s) of the 
transition) of the pre-condition are obtained. 
If these states do not exist in the global state, and cannot be
inferred therefrom, then they are added to the global state.

\item 
Get the stimulus of the pre-condition of the inspected stimulus,
call it $newStimulus$.
If $newStimulus$ is not external ($Ext$), then set the inspected
stimulus to the $newStimulus$, and go back to step 2.
\end{enumerate}

The second step considers post-conditions and adds system components 
that will be affected by the stimulus.
While the third and forth steps synthesize the components necessary to
trigger the stimulus. 
The procedure given in Appendix III 
synthesizes minimum 
topologies necessary to trigger a given stimulus of the protocol. 

Note that there may be several pre-conditions or post-conditions
for a 
stimulus, in which case several choices can be made. These
represent 
branching points in the search space.
At the end of this stage, the global state to be investigated is 
obtained.

\subsubsection{Forward Implication}
~\\
The states following $G_I$ (i.e. $G_{I+i}$ where $i > 0$) are
obtained 
through forward implication. 
We simply apply the transitions, starting from $G_I$, as given by
the transition table, in addition to implied transitions (such as
timer implication).
Furthermore, faults are incorporated into the search. For example,
in the case of a message loss, the transition 
that would have resulted from the message is not applied.
If more than one state is affected by the message, then the
space
is expanded to include the various selective loss scenarios for
the affected routers.
For crashes, the routers affected by the crash transit into the 
crashed state as defined by the expanded transition rules, as will
be shown in Section~\ref{fotg_apply}.
Forward implication uses the forward search techniques described earlier
in Section~\ref{forward}.

According to the transition completion concept (see 
Section~\ref{transition_completion}), the proper analysis of behavior 
should start from externally triggered transitions.
For example, the analysis should not consider a $Join$ without
considering the $Prune$ triggering it and its effects on the system. 
Thus, the global system state 
must be rolled back to the beginning of a complete transition
(i.e. the previous stable state) before applying the forward implication.
This will be implied in the forward implication algorithm to simplify the 
discussion.

\subsubsection{Backward Implication}
~\\
Backward implication attempts to obtain a 
sequence of events leading to $G_I$, from an initial state
($I.S.$), if 
such a sequence exists; i.e. if $G_I$ is reachable from $I.S.$

The state dependency table described in Section~\ref{dep_table} 
is used in the backward search.

Backward steps are taken for the components in the  global state
$G_I$, each step producing another global state $GState$.
For each state in $GState$ possible backward
implication rules are attempted to obtain valid backward steps toward
an initial state.
This process is repeated for preceding states in a depth first 
fashion.
A set of visited states is maintained to avoid looping. 
If all backward branches are exhausted and no initial state was reached 
the state is declared unreachable.

To rewind the global state one step backward, the reverse transition 
rules are applied. Depending on the stimulus type
of the backward rule, different states in $GState$ are
rolled back. For $orig$ and $dst$ only the originator and
destination of the stimulus is rolled back, respectively.
For $mcast$, all affected states are rolled back except
the originator. $mcastDownstream$ is similar to $mcast$
except that all downstream routers or states are rolled back,
while only one upstream router (the destination) is rolled back.
Appendix III 
shows procedures `Backward' and `Rewind' that 
implement the above steps.

Note, however, that not all backward steps are valid, and backtracking is
performed when a backward step is invalid.
Backtracking may occur when the preceding states contradict the
rules of the protocol. These contradictions may manifest themselves as: 

\begin{itemize}
\item $Src$ not found: $src$ is the originator of the stimulus,
and the global state has to include at least one component to
originate the stimulus. An example of this contradiction occurs
for the $Prune$ stimulus, for a global state $\{NH,F,NF\}$, where
the an originating component of the $Prune$ ($NC$ in this case)
does not belong to the global state.
\item Failure of minimum topology check: the necessary conditions
to trigger the stimulus must be present in the global topology.
Examples of failing the minimum topology check include, for
instance, $Join$ stimulus with global state $\{NH, NF\}$, or
$Assert$ stimulus with global state $\{F,NH,NC\}$.
\item Failure of consistency check: to maintain consistency of
the transition rules in the reverse direction, we must check that
every backward step has an equivalent forward step. To achieve
this, we must check that there is no transition $x \rightarrow y$
for the given stimulus, such that $x \in GState$. Since if $x$
remains in the preceding global state, the corresponding forward
step would transform $x$ into $y$ and the system would exist in a
state inconsistent with the initial global state (before the
backward step). An example of this inconsistency exists when the
stimulus is $FPkt$ and $GState = \{F,NF,EU\}$, where $EU
\rightarrow F$ is a post condition for $FPkt$.
See Appendix III 
for the consistency check procedure.
\end{itemize}

\subsection{Applying The Method}
\label{fotg_apply}

In this section we discuss how the fault-oriented test 
generation can be applied to the model of PIM-DM. 
Specifically, we discuss in details the application of FOTG to the 
robustness analysis of PIM-DM in the presence of single message 
loss and machine crashes.
We first walk through a simple illustrative example. 
Then we present the results of the case study in terms
of correctness violations captured by the method.

\subsubsection{Method input}
\label{fotg_input}

The protocol model is provided by the designer or protocol specification, 
in terms of a transition table~\footnote{The traditional input/output 
transition table is sufficient for our method. The pre/post-condition 
transition table can be derived automatically therefrom.}, and the 
semantics of the messages. In addition, a list of faults to be studied is 
given as input to the method. For example, definition of the fault as 
single selective protocol message loss, applied to the list of messages 
$\{Join, Prune, Assert, Graft\}$. Also a set of initial state symbols, in 
our case $\{NM, EU\}$. A definition of the design requirement, in 
this case definition of correctness, is also provided by the 
specification. The rest of the process is automated.

\subsubsection{Illustrative example}
\label{fotg_example}

Figure~\ref{topo_construct} shows the phases of FOTG for a simple example
of a $Join$ loss. Following are the steps taken for that example:

\begin{figure}[t] 
 \begin{center}
  \epsfig{file=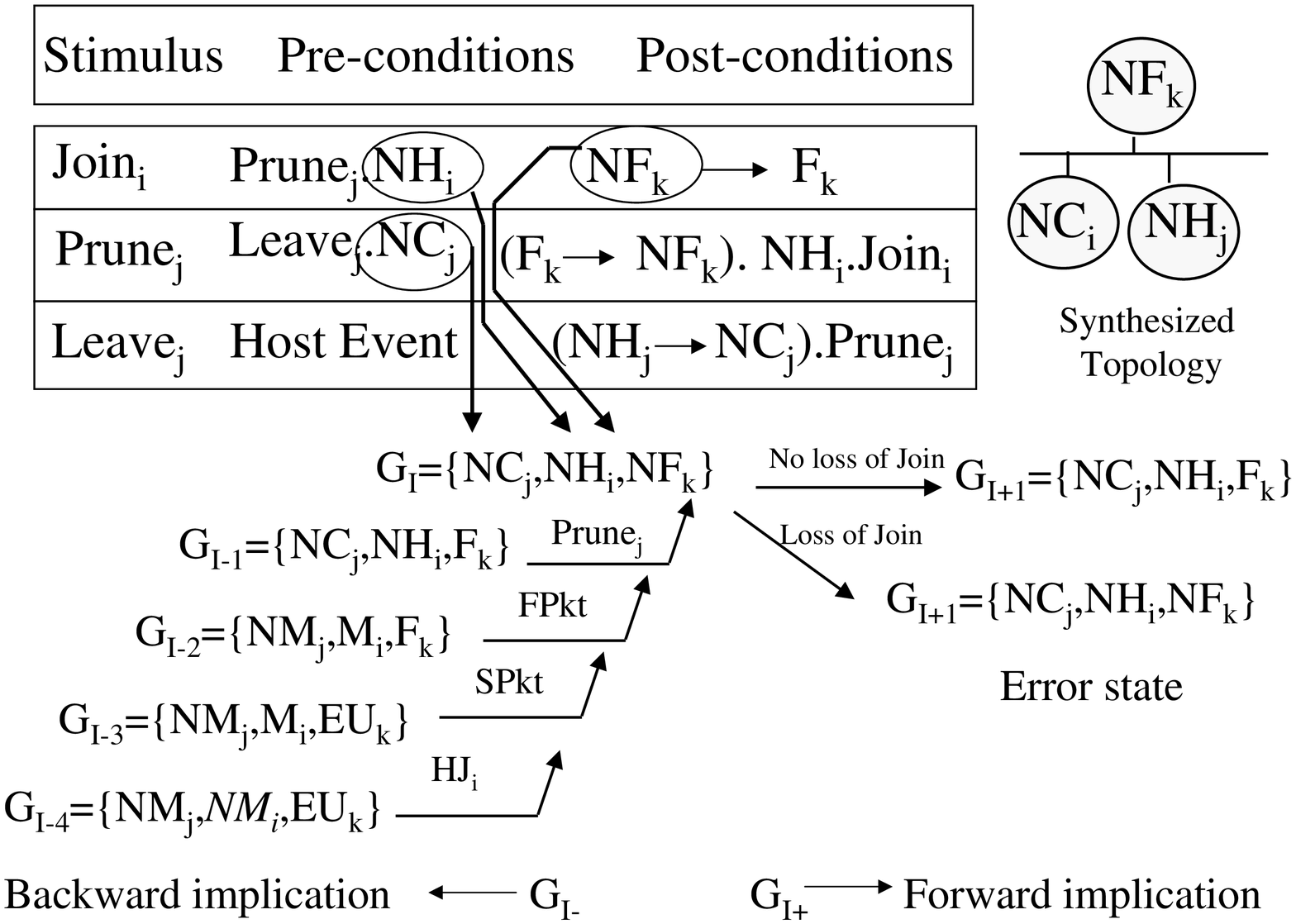,height=7.5cm,width=7.5cm,angle=0}
   \caption{$Join$ topology synthesis, forward/backward 
implication}\label{topo_construct}
 \end{center}
\end{figure}

\scriptsize

\hspace{-.6cm}
\begin{tabular}{p{3.4in}} \hline
{\bf Synthesizing the Global State} \\ \hline
1. $Join$: $startState$ of post-condition is $NF_{dst} 
\Rightarrow G_I = \{NF_{k}\}$ \\ \hline
2. $Join$: state of pre-condition is $NH_i \Rightarrow G_I =
\{NH_i,NF_{k}\}$, goto $Prune$ \\ \hline
3. $Prune$: $startState$ of post-condition is $F_k$, implied from $NF_{k}$ 
in $G_I$ \\ \hline
4. $Prune$: state of pre-condition is $NC_j \Rightarrow
G_I = \{NH_i,NF_{k}, NC_j\}$, goto $L$ (Ext) \\ \hline
5. $startState$ of post-condition is $NH$ can be implied
from $NC$ in $G_I$ \\ \hline
\end{tabular}

\hspace{-.6cm}
\begin{tabular}{p{3.4in}} \hline
{\bf Forward implication} \\ \hline
without loss: $G_I = \{NH_i,NF_{k},NC_j\} 
\stackrel{Join}{\longrightarrow}
G_{I+1} = \{NH_i,F_k,NC_j\}$ \\ \hline
loss w.r.t. $R_j$:
$\{NH_i,NF_{k},NC_j\} \longrightarrow
G_{I+1} = \{NH_i,NF_k,NC_j\}$ error \\ \hline
\end{tabular}

\hspace{-.6cm}
\begin{tabular}{p{3.4in}} \hline
{\bf Backward implication} \\ \hline
$G_I = \{NH_i,NF_{k},NC_j\} 
\stackrel{Prune}{\longleftarrow}
G_{I-1} =
\{NH_i,F_k,NC_j\} 
\stackrel{FPkt}{\longleftarrow}
G_{I-2} =$ \\ 
$\{M_i,F_k,NM_j\} 
\stackrel{SPkt}{\longleftarrow}
G_{I-3} = \{M_i,EU_k,NM_j\}
\stackrel{HJ_i}{\longleftarrow}
G_{I-4} = \{NM_i,EU_k,NM_j\} = I.S.$ \\ \hline
\end{tabular}

\normalsize

Losing the $Join$ by the forwarding router $R_k$ leads to an error state 
where router $R_i$ is expecting packets from the LAN, but the LAN has no 
forwarder. 

\subsubsection{Summary of Results}
\label{fotg_analysis}

In this section we briefly discuss the results of applying our method to 
PIM-DM.
The analysis is conducted for single message loss and momentary loss of 
state. For a detailed analysis of the results see
Appendix III-G. 

\paragraph{\bf Single message loss}

We have studied single message loss scenarios for the $Join, Prune, Assert,$ 
and $Graft$ messages.
For this subsection, we mostly consider non-interleaved external events, 
where the system is stimulated only once between stable states.
The $Graft$ message is particularly interesting, since it is acknowledged, 
and it raises timing and sequencing issues that we address in a later 
subsection, where we extend our method to consider interleaving of 
external events.

Our method as presented here, however, may not be generalized to 
transform any type of timing problem into sequencing problem. This topic 
bears more research in the future.

We have used the sequences of events generated automatically by the algorithm
to analyze protocol errors and suggest fixes for those errors.

{\bf Join:} A scenario similar to that presented in 
Section~\ref{fotg_example} incurred an error.
In this case, the robustness violation was not allowing another chance to 
the downstream router to send a $Join$. A suggested fix would be to send 
another prune by $F_{Del}$ before the timer expires.

{\bf Prune:} In the topology above, an error occurs when $R_i$ 
loses the $Prune$, hence no $Join$ is triggered. The fix suggested above
takes care of this case too.

{\bf Assert:} An error in the $Assert$ case occurs with no downstream
routers; e.g. $G_I = \{F_i,F_j\}$. The design error is the absence of a
mechanism
to prevent pruning packets in this case. One suggested fix would be to have
the $Assert$ winner schedule a deletion timer (i.e. becomes $F_{Del}$) and have 
the downstream receiver (if any) send $Join$ to the $Assert$ winner.


\begin{figure}[t]
 \begin{center}
  \epsfig{file=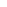,height=7.5cm,width=5cm,angle=270,clip=}
   \caption{Graft event sequencing}\label{graft_seq}
 \end{center}
\end{figure}

{\bf Graft:} 
A $Graft$ message is acknowledged by $GAck$, hence the protocol did not incur
error when the $Graft$ message was lost
with non-interleaved external events.
The protocol is robust to $Graft$ loss with the use of $Rtx$ timer.
Adversary external conditions are interleaved during the transient states and
the $Rtx$ timer is cleared, such that the adverse event will not be 
overridden by the $Rtx$ mechanism. 

To clear the $Rtx$ timer, a transition should be created from 
$NH_{Rtx}$ to $NH$ which is triggered by a $GAck$ according to the state 
dependency table ($NH 
\stackrel{GAck}{\longleftarrow}
NH_{Rtx}$).
This transition is then inserted in the event sequence, and forward and
backward implications are used to obtain the overall sequence of events 
illustrated in Figure~\ref{graft_seq}.
In the first and second scenarios ({\bf I} and {\bf II}) no error occurs.
In the third scenario ({\bf III}) when a $Graft$ followed by a $Prune$ is 
interleaved with the $Graft$ loss, the $Rtx$ timer is reset with 
the receipt of the $GAck$ for the first $Graft$, and the systems ends up 
in an error state. A suggested fix is to add sequence numbers to
$Graft$s, at the expense of added complexity.

\paragraph{\bf Loss of State}
\label{loss_of_state}

We consider momentary loss of state in a router.
A `$Crash$' stimulus transfers the crashed router from any state 
`X' into `EU' or `ED'.
Hence, we add the following line to the transition table:

\footnotesize

\begin{tabular}{lll} \hline
{\bf Stimulus} & {\bf Pre-cond} & {\bf
Post-cond} (stimulus.state/trans) \\ \hline 
$Crash$ & Ext & $\{NM,M,NH,NC,NH_{Rtx}\} \rightarrow ED$, \\
& & $\{F,F_{Del},NF\} \rightarrow EU$ \\ \hline
\end{tabular}

\normalsize

The FSM resumes function immediately after 
the crash (i.e. further transitions are not affected).
We analyze the 
behavior when the crash occurs in any router state.
For every state, a topology is synthesized that is necessary to create 
that state.
We leverage the topologies previously synthesized for the messages. 
For example, state $F_{Del}$ may be created from state $F$ by receiving a 
$Prune$ ($F_{Del} 
\stackrel{Prune}{\longleftarrow}
F$). 
Hence we may use the topologies constructed for $Prune$ loss to analyze 
a crash for $F_{Del}$ state.

Forward implication is then applied, and behavior 
after the crash is checked for correct packet delivery.
To achieve this, host stimuli (i.e. $SPkt$, $HJ$ and $L$) are applied,
then the system state is checked for correctness.

In lots of the cases studied, the system recovered from the crash (i.e. 
the system state was eventually correct). 
The recovery is mainly due to the nature of PIM-DM; where protocol states 
are re-created with reception of data packets.
This result is not likely to extend to protocols
of other natures; e.g. PIM Sparse-Mode~\cite{PIM-SMv2-Spec}.

\begin{figure}[th]
 \begin{center}
  \epsfig{file=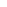,height=8.5cm,width=3cm,angle=270,clip=}
   \caption{Crash leading to join latency}\label{crash_bad_1} 
 \end{center}
\end{figure}

However, in violation with robustness requirements, there existed
cases in which the system did not recover.
In Figure~\ref{crash_bad_1}, the 
host joining in (II, a) did not have the sufficient state to send a 
$Graft$ and hence gets join latency until the negative cache state 
times out upstream and packets are forwarded onto the LAN as in (II, b).

\begin{figure}[th]
 \begin{center}
  \epsfig{file=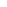,height=8.5cm,width=3cm,angle=270,clip=}
   \caption{Crash leading to black holes}\label{crash_bad_2} 
 \end{center}
\end{figure}

In Figure~\ref{crash_bad_2} (II, a), the 
downstream router incurs join latency due to the crash of the upstream 
router. The state is not corrected until the periodic broadcast 
takes place, and packets are forwarded onto the LAN as in (II, b).

\subsection{Challenges and Limitations}
\label{fotg_discuss}

Although we have been able to apply FOTG to PIM-DM successfully,
a discussion of the open issues and challenges is called for.
In this section we address some of these issues.

\begin{itemize}

\item The topologies synthesized by the above FOTG study are only limited to a
single-hop LAN with $n$
routers~\footnote{This limitation is similar to that suffered by FITG in
Section~\ref{forward}.}. This means that the above FOTG analysis is necessary but not
sufficient to verify robustness of the end-to-end behavior of the 
protocol in a multi-hop topology; even if each LAN
in the topology operates correctly, the inter-LAN interaction may introduce
erroneous behaviors. 
Applying FOTG to multi-hop topologies is part of future research.

\item The analysis for our case studies did not consider network delays. In order to
study end-to-end protocols network delays must be considered in the model.
In~\cite{e2e} we introduce the notion of $virtual$ LAN to include
end-to-end delay semantics.

\item Minimal topologies that are necessary and sufficient to trigger the
stimuli, may not be sufficient to capture all correctness violations. For
example, in some cases it may require one member to trigger a $Join$, but two
members to experience an error caused by $Join$ loss. Hence, the topology
synthesis stage must be complete in order to capture all possible
errors. To achieve this we propose to use the symbolic representation. For 
example, to cover all topologies with one or more members we use 
($M^{1+}$). Integration of this notation with the full method is part of 
future work.

\item The efficiency of the backward search may be increased using reduction
techniques, such as equivalence of states and transitions (similar to the ones
presented in Section~\ref{forward}). 
In addition, the algorithm complexity may be reduced by utilizing
information about reachable states to reduce the search. This information
could be obtained simply by storing previous sequences and
states visited. Alternatively, the designer may provide information --based
on protocol-specific knowledge-- about reachable states, through a compact
representation thereof.

\item The topologies constructed by FOTG are inferred from the mechanisms 
specified by the transition table of the GFSM. The FOTG algorithm will 
not construct topologies resulting from non-specified mechanisms. For 
example, if the $Assert$ mechanism that deals with duplicates was left out (due
to a design error) the algorithm would not construct $\{F_i,F_j\}$ topology.
Hence, FOTG is not guaranteed to detect duplicates in this case.
So, FOTG (as presented here) may be used to evaluate behavior of specified
mechanisms in the presence of network failures, but is not a general 
protocol verification tool.

\item The global states synthesized during the topology synthesis phase are not
guaranteed to be reachable from an initial state. Hence the algorithm may be
investigating non-reachable states, until they are detected as unreachable in
the last backward search phase.
	Adding reachability detection in the early stages of FOTG is subject of
future work. However, statistics collected in our case study (see 
Appendix III-F) 
show that unreachable states are not the 
determining factor in the complexity of the backward search. Hence, 
other reduction techniques may be needed to increase the efficiency of 
the method.

\end{itemize}

We believe that the strength of our fault-oriented method, as was 
demonstrated, lies in its ability to construct the necessary conditions 
for erroneous behavior by starting directly from the fault and avoiding 
the exhaustive walk of the state space. Also, converting timing problems 
into sequencing problems (as was shown for $Graft$ analysis) reduces the 
complexity required to study timers.
FOTG as presented in this chapter seems best fit to study protocol 
robustness in the presence of faults. Faults presented in our studies 
include single selective loss of protocol messages and router crashes.


\section{Related Work}
\label{related}

The related work falls mainly in the field of 
protocol verification, distributed algorithms
and conformance testing.
In addition, some concepts of our work were inspired by VLSI chip testing.
Most of the literature on multicast protocol design addresses 
architecture, specification, and comparisons between different protocols. 
We are not aware of any other work to develop systematic methods for 
test generation for multicast protocols.

There is a large body of literature dealing with verification of 
communication protocols.
Protocol verification is the problem of ensuring the logical consistency 
of the protocol specification, independent of any particular
implementation.
Protocol verification typically addresses well-defined properties, such 
as safety (e.g., freedom from deadlocks) and liveness (e.g., absence of
non-progress cycles)~\cite{proto_design}.
In general, the two main approaches for protocol verification are 
theorem proving and reachability analysis (or model 
checking)~\cite{formal_survey1}~\cite{formal_survey2}.
In theorem proving, system properties are expressed in logic formulas,
defining a set of axioms and constructing relations on these axioms. 
In contrast to reachability analysis, theorem proving can deal with infinite 
state spaces.
Interactive theorem provers require human intervention, and
hence are slow and error-prone.
Theorem proving includes {\em model-based} and {\em logic-based} formalisms.
Model-based formalisms (e.g., Z~\cite{z}, VDM~\cite{vdm}) are suitable for protocol
specifications in a succinct manner, but lack the tool support for
effective proof of properties.
The use of first order logic allows the use of theorem provers (e.g.,
Nqthm~\cite{nqthm}), but may result in 
specifications that are difficult to read.
Higher order logic (e.g., PVS~\cite{pvs}) provides expressive power for
clear descriptions and proof capabilities for protocol properties.
The number of axioms and relations grows with the complexity of the protocol.
Axiomatization and proofs depend largely on human intelligence, which limits the use
of theorem proving systems. Moreover, these systems tend to abstract out network
failures we are addressing in this study.

Reachability analysis algorithms~\cite{reachability}~\cite{reachability2}
attempt to generate and inspect all the protocol states that are 
reachable from given initial states. 
The main types of reachability analysis algorithms include full  
search and controlled partial search.
If full search exceeds the memory or time limits, it effectively reduces
to an uncontrolled partial search, and the quality of the analysis   
deteriorates quickly.
Such algorithm suffers from the 
`state space explosion' problem, especially for complex protocols.
To circumvent this problem, state reduction and controlled partial search 
techniques~\cite{partial_reachability}~\cite{partial_reachability2} could
be used. 
These techniques focus only on parts of the state space and may use 
probabilistic~\cite{prob_reachability}, random~\cite{random_reachability} or 
guided searches~\cite{guided_reachability}.
In our work we adopt approaches extending reachability
analysis for multicast protocols. Our fault-independent test
generation method (in Section~\ref{forward}) 
borrows from controlled partial search and state reduction techniques.

Work on distributed algorithms deals with synchronous networks, 
asynchronous shared memory and asynchronous networked systems~\cite{lynch}.
Proofs can be established using an automata-theoretic framework. 
Several studies on distributed algorithms
considered failure models including message loss or duplication,
and processor failures, such as stop (or crash) failures,
transient failures, or byzantine failures~\cite{lamport}, where
failed processors behave arbitrarily.
We do not consider byzantine failures in our study.
Distributed algorithms may be treated in a formal
framework, using automata-theoretic models and state machines,
where results are presented in terms of set-theoretic
mathematics~\cite{lynch}. The formal framework is used to present proofs or
impossibility results.
Proof methods for distributed algorithms 
include invariant assertions and simulation relationships~\footnote{An invariant
assertion is a property that holds true for all
reachable states of the system, while a simulation is a
formal relation between an abstract solution of the problem and a
detailed solution.} that
are generally proved using induction, and may be checkable
using theorem-provers, e.g., Larch theorem-prover~\cite{larch}.
Asynchronous network components can be modeled as 
timed-automata~\cite{timed_automata},~\cite{lynch}.

Several attempts to apply formal verification to network protocols have been 
made. Assertional proof techniques were used to prove distance 
vector routing~\cite{Taji77:Correctness}, path vector 
routing~\cite{Shin87:Performance} and route diffusion 
algorithms~\cite{Jaff82:Responsive,Garc93:Loop-Free} 
and~\cite{Merl79:Failsafe} using communicating finite state machines.
An example point-to-point mobile application was proved using assertional
reasoning in~\cite{Roma96:Assertional} using UNITY~\cite{Chan88:Parallel}.
Axiomatic reasoning was used in proving a simple transmission 
protocol in~\cite{Hail85:Simple}.
Algebraic systems based on the calculus of communicating systems 
(CCS)~\cite{Miln80:Calculus} have been used to prove 
CSMA/CD~\cite{Parr88:Verifying}.
Formal verification has been applied to TCP and T/TCP in~\cite{ttcp}.

Multicast protocols may be modeled as asynchronous networks, with the components as
timed-automata, including failure models. In fact, the global
finite state machine (GFSM) model used by our search
algorithms is adopted from asynchronous shared memory systems (in
specific, cache coherence algorithms~\cite{cache_coherence}) and
extended with various multicast and timing semantics.
The transitions of the I/O automaton may be given in the form of
pre-conditions and effects~\footnote{This is similar to our representation
of the transition table for the fault-oriented test generation
method.}.

The combination of timed automata, invariants, simulation mappings,
automaton composition, and temporal logic~\cite{temporal_logic} seem to be 
very useful tools for proving (or disproving) and reasoning about safety or
liveness properties of distributed algorithms.
It may also be used to establish asymptotic bounds on
the complexity of the distributed algorithms.
It is not clear, however, how theorem proving techniques can be
used in test synthesis to construct event sequences and
topologies that stress network protocols. 
Parts of our work draw from distributed
algorithms verification principles. Yet we feel that our work
complements such work, as we focus on test synthesis problems. 

Conformance Testing is used to check that the external behavior of a given
implementation of a protocol is equivalent to its formal specification.
A conformance test fails if the implementation and specification
differ. By contrast, verification of the protocol must always reveal the design 
error. Given an implementation under test (IUT), sequences of input messages are 
provided and the resulting output is observed. The test passes only
if all observed outputs matche those of the formal specification.
The sequences of input messages is called a conformance test suite and the main
problem is to find an efficient procedure for generating a
conformance test suite for a given protocol. One possible solution is to generate a
sequence of state transitions that passes through every state and every transition at
least once; also known as a transition tour~\cite{klee}.
The state of the machine must be checked after each transition with the help of 
unique input/output (UIO)
sequences~\footnote{A Unique Input/Output (UIO) sequence is a sequence of
transitions that can be used to determine the state of the IUT.}.
To be able to verify every state in the IUT, we must be able to derive a 
UIO sequence for every state separately.
This approach generally suffers from the following drawbacks.
Not all states of an FSM have a UIO sequence.
Even if all states in a FSM have a UIO sequence, the problem of deriving
UIO sequences has been proved to be p-complete in~\cite{pcomplete}; 
i.e. only very short UIO sequences can be found in
practice~\footnote{In~\cite{yanalee} a
randomized polynomial time algorithm is presented for designing UIO checking
sequences.}. UIO sequences can identify states
reliably only in a correct IUT. Their behavior for faulty IUTs is unpredictable, and
they cannot guarantee that any type of fault in an IUT remains detectable.
Only the presence of desirable behavior can be tested by 
conformance testing, not the absence of undesirable behavior.

Conformance testing techniques are important for testing 
protocol implementations.
However, it does not target design errors or protocol performance.
We consider work in this area as complementary to the focus of 
our study.

VLSI Chip testing uses a set of well-established approaches to generate test 
vector patterns, generally for detecting physical defects in the VLSI 
fabrication process.
Common test vector generation methods detect single-stuck faults; where 
the value of a line in the circuit is always at logic `1' or `0'.
Test vectors are generated based on a model of the circuit and a given 
fault model. 
Test vector generation can be fault-independent or
fault-oriented~\cite{testability}~\cite{bist}.
In the fault-oriented process, the two fundamental steps in generating a 
test vector are to activate (or excite) the fault, and to propagate 
the resulting error to an observable output.
Fault excitation and error propagation usually involve a search 
procedure with a backtracking strategy to resolve or undo contradiction 
in the assignment of line and input values.
The line assignments performed sometimes determine or imply other line 
assignments.
The process of computing the line values to be consistent with previously 
determined values is referred to as {\em implication}.
Forward implication is implying values of lines from the fault toward the 
output, while backward implication is implying values of lines from the 
fault toward the circuit input.
Our approaches for protocol testing use some of the above principles; 
such as forward and backward implication.
VLSI chip testing, however, is performed
a given circuit, whereas protocol testing is performed for arbitrary
and time varying topologies.

Other related work includes verification of cache 
coherence protocols~\cite{cache_coherence}. This study
uses counting equivalence relations and symbolic representation of states to reduce
space search complexity. We use the notion of counting equivalence in our study.

\section{Conclusions}
\label{conclusion}

In this study we have proposed the {\em STRESS} framework to integrate test
generation into the protocol design process. Specifically, we targeted automatic test
generation for robustness studies of multicast routing protocols. We have adopted a
global FSM model to represent the multicast protocols on a LAN. In addition, we have
used a fault model to represent packet loss and machine crashes. We have investigated
two algorithms for test generation; namely, the fault-independent test generation
(FITG) and the fault-oriented test generation (FOTG). Both algorithms were used to
study a standard multicast routing protocol, PIM-DM, and were compared in terms of
error coverage and algorithmic complexity. For FITG, equivalence reduction
techniques were combined with forward search to reduce search complexity from exponential to
polynomial. FITG does not provide topology synthesis.
For FOTG, a mix of forward and backward search techniques allowed for automatic synthesis
of the topology. We believe that FOTG is a better fit for robustness studies since it
targets faults directly. The complexity for FOTG was quite manageable for our case
study. Corrections to errors captured in the study were proposed with the
aid of our method and integrated into the latest PIM-DM specification. More case
studies are needed to show more general applicability of our methodology.

\appendix

\section{State Space Complexity}
\label{search_apndx}


In this appendix we present analysis for the state space complexity of 
our target system. In specific we present completeness proof of the state 
space and the formulae to compute the size of the correct state space.

\subsection{State Space Completeness}
\label{completeness}

We define the space of all states as $X^*$, denoting zero or more
routers in any state. We also define the algebraic operators for
the space, where 
\begin{equation}
X^* = X^0 \cup X^1 \cup X^{2+}
\end{equation}
\begin{equation}
\left( Y^n, X^* \right) = \left( Y^{n+}, \{X-Y\}^* \right)
\end{equation}

\subsubsection{Error states}

In general, an error may manifest itself as packet duplicates,
packet loss, or wasted bandwidth. This is mapped onto the state
of the global FSM as follows:

\begin{enumerate}
\item The existence of two or more forwarders on the LAN with one
or
more routers expecting packet from the LAN (e.g., in the $NH_X$
state) indicates duplicate delivery of packets.

\item The existence of one or more routers expecting packets from
the LAN with no forwarders on the LAN indicates a deficiency in
packet delivery (join latency or black holes).

\item The existence of one or more forwarders for the LAN with no
routers expecting packets from the LAN indicates wasted
bandwidth (leave latency or extra overhead).
\end{enumerate}

- for duplicates: one or more $NH_X$ with two or more $F_X$;  

\begin{equation}
\left( NH_X, F_X^{2+}, X^* \right)
\end{equation}

- for extra bandwidth: one or more $F_X$ with zero $NH_X$;

\begin{equation}
\left( F_X, \{ X - NH_X \}^* \right)
\end{equation}

- for blackholes or packet loss: one or more $NH_X$ with zero
$F_X$;
\begin{equation}
\left( NH_X , \{ X - F_X \}^* \right)
\end{equation}

\subsubsection{Correct states}

As described earlier, 
the correct states
can be described by the following rule:

{\em $\exists$ exactly one forwarder for the LAN iff $\exists$ 
one or more routers expecting packets from the LAN}.

- zero $NH_X$ with zero $F_X$; 

\begin{equation}
\left( \{X - NH_X - F_X\}^* \right)
\end{equation}

- one or more $NH_X$ with exactly one $F_X$; 

\begin{equation}
\left( NH_X, F_X, \{ X - F_X \}^* \right)
\end{equation}

from (B.2) and (B.3) we get:

\begin{equation}
\left( NH_X, F_X^{2+}, \{ X - F_X \}^* \right)
\end{equation}

if we take the union of (B.8), (B.5) and (B.7), and apply (B.1) we get:

\begin{equation}
\left( NH_X, X^* \right) = \left( NH_X^{1+}, \{ X - NH_X \}^* \right)
\end{equation}

also, from (B.4) and (B.2) we get:

\begin{equation}
\left( F_X^{1+}, \{ X - NH_X - F_X \}^* \right)
\end{equation}

if we take the union of (B.10) and (B.6) we get:

\begin{equation}
\left( F_X^*, \{ X - NH_X - F_X \}^* \right) = \left( \{ X - NH_X \}^* \right)
\end{equation}

taking the union of (B.9) and (B.11) we get:

\begin{equation}
\left( NH_X^*, \{ X - NH_X \}^* \right) = \left( X^* \right)
\end{equation}

which is the complete state space.

\subsection{Number of Correct and Error State Spaces}
\label{numbers}

\subsubsection{First case definition}
\label{first_number}

For the correct states:
$\left( \{X-NH-F\}^* \right)$ reduces the symbols from which to
choose the state by 2; i.e. yields the formula:

\[ C(n+(s-2)-1,n) = C(n+s-3,n). \] 

While $\left( NH, F, \{X-F\}^* \right)$
reduces the number of routers to choose by 2 and the number of
symbols by 1, yielding: 

\[ C((n-2)+(s-1)-1, n-2) = C(n+s-4,n-2). \]

\subsubsection{Second case definition}
\label{second_number}

For the correct states:
$\left( \{X-NH_X-F_X\}^* \right)$ reduces, the number of states
by 4, yielding 

\[ C(n+(s-4)-1,n) = C(n+s-5,n). \] 

While $\left( NH_X, F_X, \{X-F_X\}^* \right)$ reduces the
number of routers to $n-2$ and the symbols to $s-2$ and yields

\[ 4 \cdot C((n-2)+(s-2)-1,n-2)) = 4 \cdot C(n+s-5,n-2). \]

We have to be careful here about overlap of sets of correct
states. For example $\left( NH, F, \{X-F_X\}^* \right)$ is
equivalent to $\left( NH_{Rtx}, F, \{X-F_X\}^* \right)$ when a
third router is in $NH_{Rtx}$ in the first set and $NH$ in the
second set. Thus we need to remove one of the sets $\left( NH,F,
NH_{Rtx}, \{X-F_X\}^* \right)$, which translates in terms of
number of states to 

\[ C((n-3)+(s-2)-1,n-3) = C(n+s-6,n-3). \] 

A similar argument is given when we replace $F$ above by
$F_{Del}$, thus we multiply the number of states to be removed by
2. Thus, we get the total number of equivalent correct states:

$C(n+s-5,n) + 4 \cdot C(n+s-5,n-2) - 2 \cdot C(n+s-6,n-3)$.

To obtain the $ErrorStates$ we can use:

\[ ErrorStates = TotalStates - CorrectStates. \]

\begin{figure}[th]
 \begin{center}
  \epsfig{file=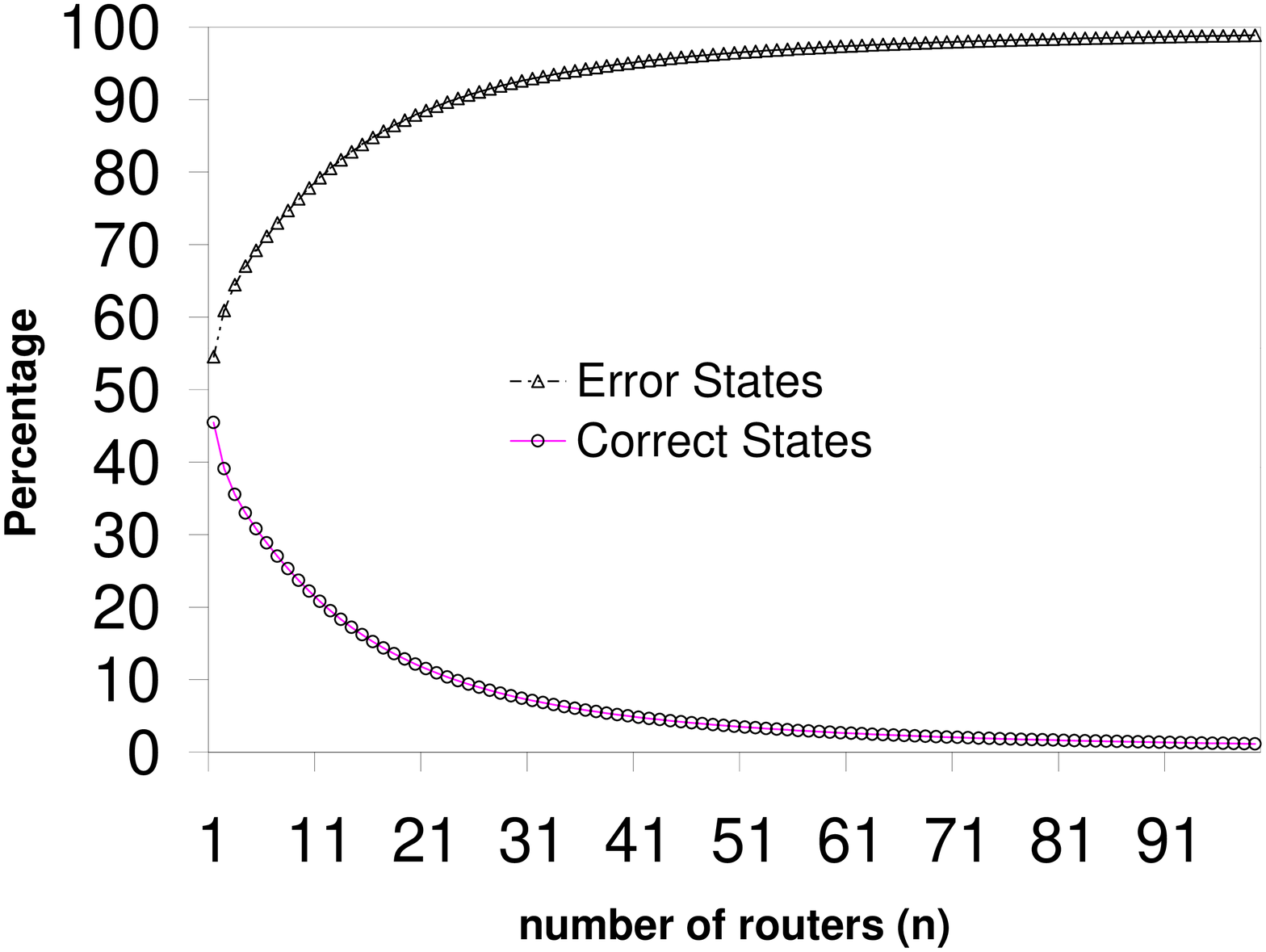,height=5cm,width=6cm,clip=,angle=0}
   \caption{The percentage of the correct and error   
states}\label{err_percentage}
 \end{center}
\end{figure}

Figure~\ref{err_percentage} shows the percentage of each of the
correct and error state spaces, and how this percentage changes
with the number of routers. The figure is shown for the second
case error definition. Similar results were obtained for the
first case definition.

\section{Forward Search Algorithms}
\label{forward_details}


This appendix includes detailed procedures that implement the forward 
search method as described in Section~\ref{forward}. It also includes 
detailed statistics collected for the case study on PIM-DM.

\subsection{Exhaustive Search}
\label{exhaustive_apndx}

The {\bf ExpandSpace} procedure given below implements an exhaustive
search, where $W$ is the working set of states to be expanded, $V$  
is the set of visited states (i.e. already expanded), and $E$ is the
state currently being explored. Initially, all the state sets
are empty. The nextState function gets
and removes the next state from $W$, according to the search
strategy; if depth first then $W$ is treated as a stack, or as a
queue if breadth first.

Each state is expanded by applying the stimuli via the `forward'
procedure that implements the transition rules and returns the 
new stable state $New$.

\scriptsize

\par \hspace{0.05in}{\bf ExpandSpace}($initGState$)\{
\par \hspace{0.1in}add $initGState$ to $W$
\par \hspace{0.1in}while $W$ not empty \{
\par \hspace{0.3in}$E$ = nextGState from $W$;
\par \hspace{0.3in}add $E$ to $V$;
\par \hspace{0.3in}$\forall$ state $\in E$
\par \hspace{0.5in}$\forall$ stim applying to state \{
\par \hspace{0.7in}$New$ = forward($E$,stim);
\par \hspace{0.7in}if $New \notin W$ or $V$
\par \hspace{0.9in}add $New$ to $W$;
\par \hspace{0.5in}\}
\par \hspace{0.1in}\}
\par \hspace{0.05in}\} \\

\normalsize

The initial state $initGState$ may be generated using the
following procedure, that produces all possible combinations of
initial states $I.S.$.

\scriptsize

\par \hspace{0.05in}{\bf Init}($depth$,$GState$)\{
\par \hspace{0.1in}$\forall state \in I.S.$ \{
\par \hspace{0.3in}add $state$ to $GState$;
\par \hspace{0.3in}$depth$ = $depth$ - 1;
\par \hspace{0.3in}if $depth$ = 0
\par \hspace{0.5in}ExpandSpace($GState$);
\par \hspace{0.3in}else
\par \hspace{0.5in}Init($depth$, $GState$);
\par \hspace{0.3in}remove last element of $GState$;
\par \hspace{0.1in}\}
\par \hspace{0.05in}\} \\

\normalsize

This procedure is called with the following parameters: (a) number
of routers $n$ as the initial $depth$ and (b) the $empty state$
as the initial $GState$.
It is a recursive procedure that does a tree search, depth first,   
with the number of levels equal to the number of routers and the    
branching factor equal to the number of initial state symbols
$|I.S.| = i.s.$.
The complexity of this procedure is given by
$(i.s.)^n$. 

\subsection{Reduction Using Equivalence}
\label{reduction_apndx}

We use the counting equivalence notion to reduce the complexity of the 
search in 3 ways:

\begin{enumerate}

\item The first reduction we use is to investigate only the   
equivalent initial states, we call this algorithm {\bf Equiv}.
One procedure that produces such equivalent initial state space
is the {\bf EquivInit} procedure given below.

\begin{figure}[t]
 \begin{center}
  \epsfig{file=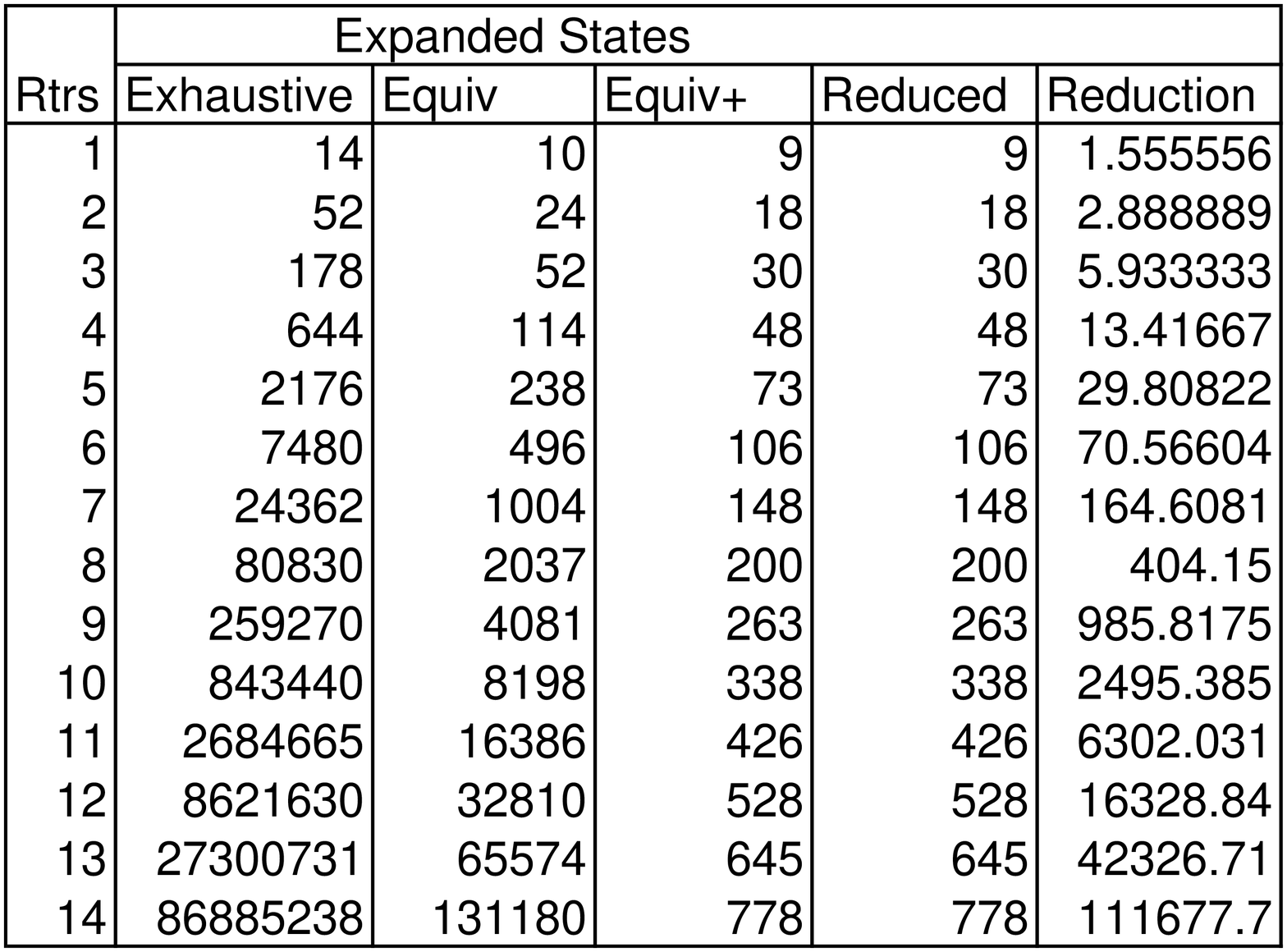,height=4cm,width=8cm,clip=,angle=0}  
   \caption{Simulation statistics for
forward algorithms. $ExpandedStates$ is the number
of visited states.}\label{tbl1}
 \end{center}
\end{figure}

\begin{figure}[t]
 \begin{center}
  \epsfig{file=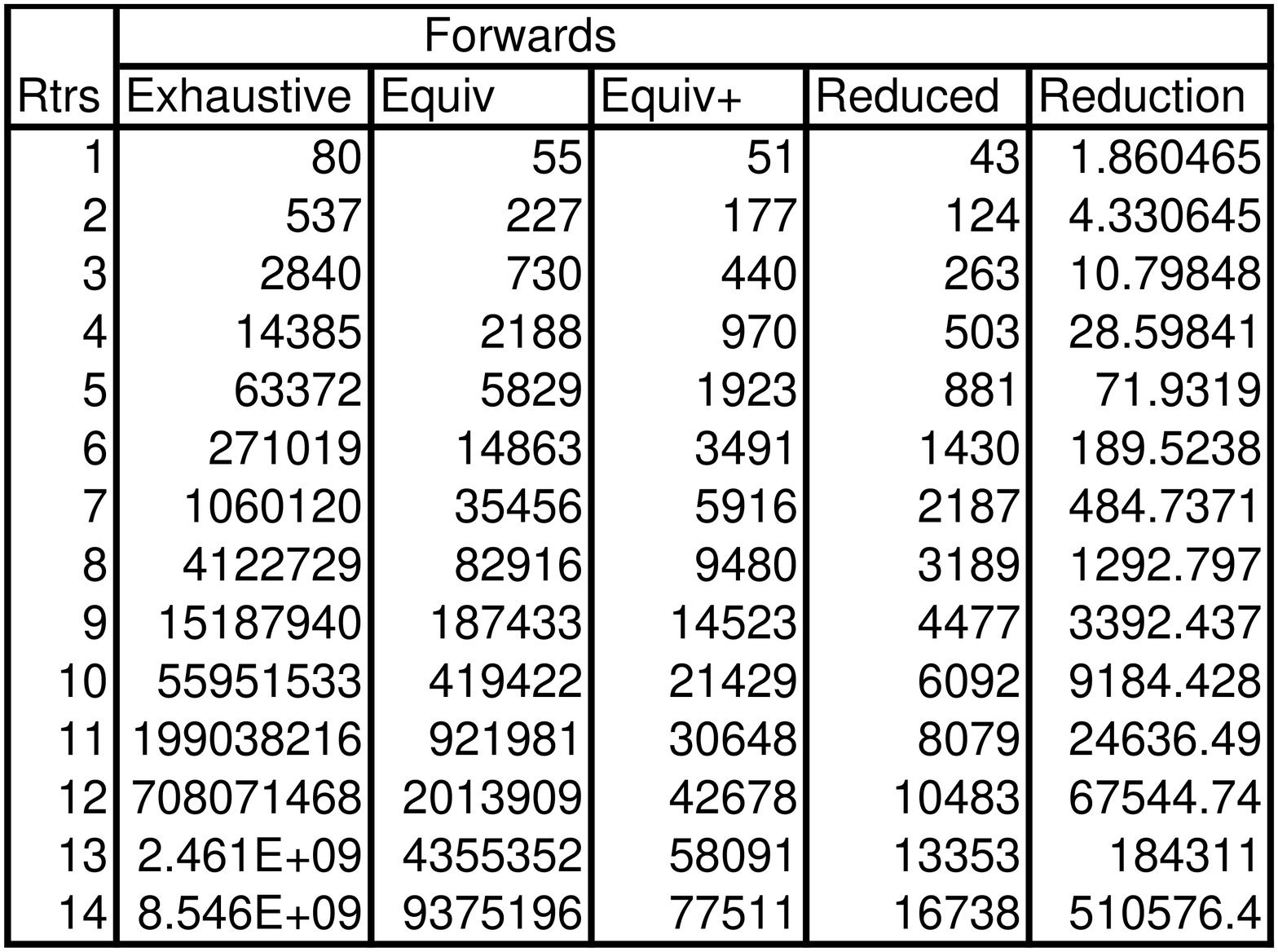,height=4cm,width=8cm,clip=,angle=0}
   \caption{Simulation statistics for
forward algorithms. $Forwards$ is the number of
calls to $forward()$.}\label{tbl1.5}
 \end{center}
\end{figure} 

\begin{figure}[t]
 \begin{center}
  \epsfig{file=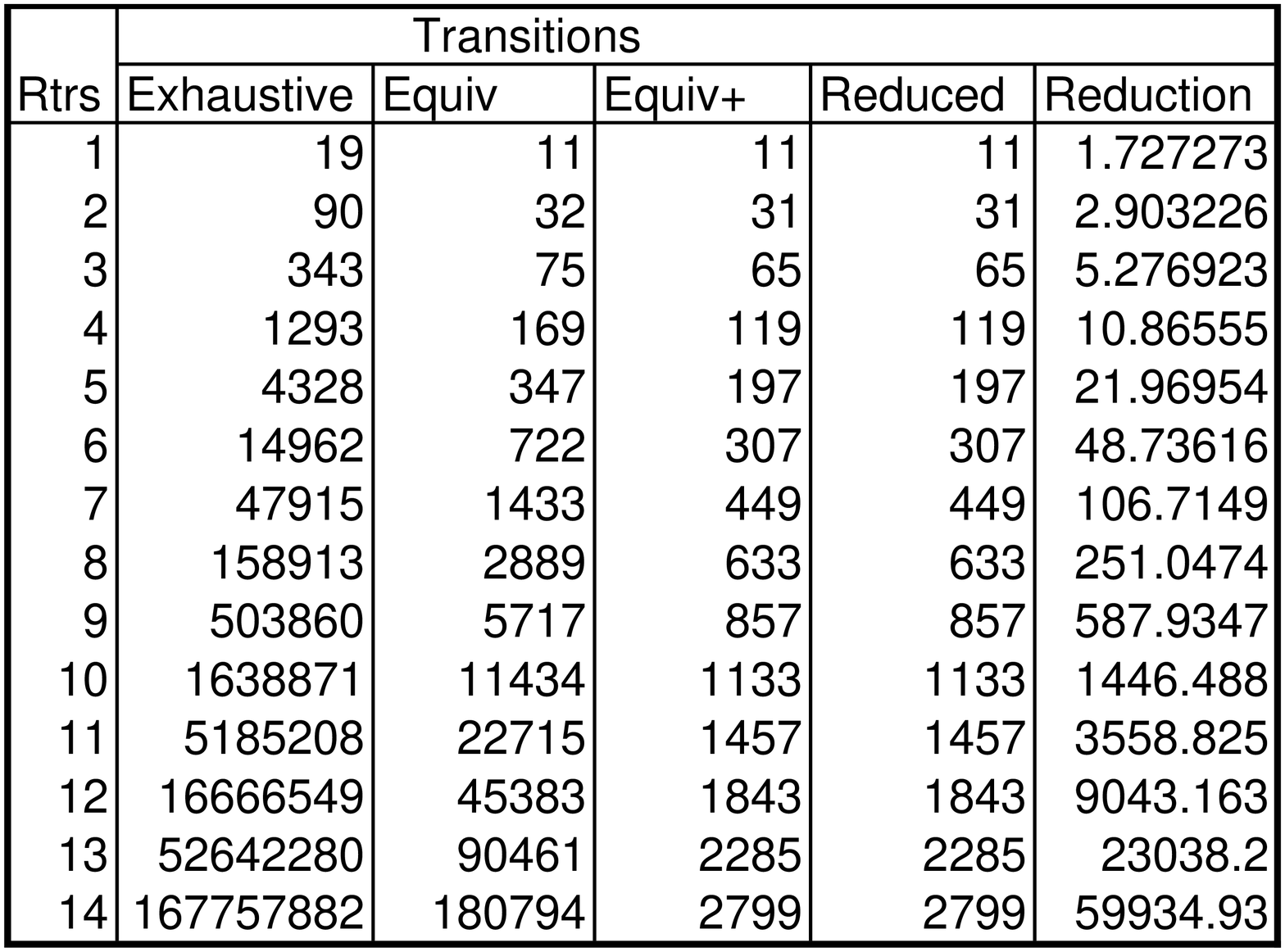,height=4cm,width=8cm,clip=,angle=0}
   \caption{Simulation statistics for
forward algorithms. $Transitions$ is the number of
transient states visited.}\label{tbl2}
 \end{center}
\end{figure}

\begin{figure}[t]
 \begin{center}
  \epsfig{file=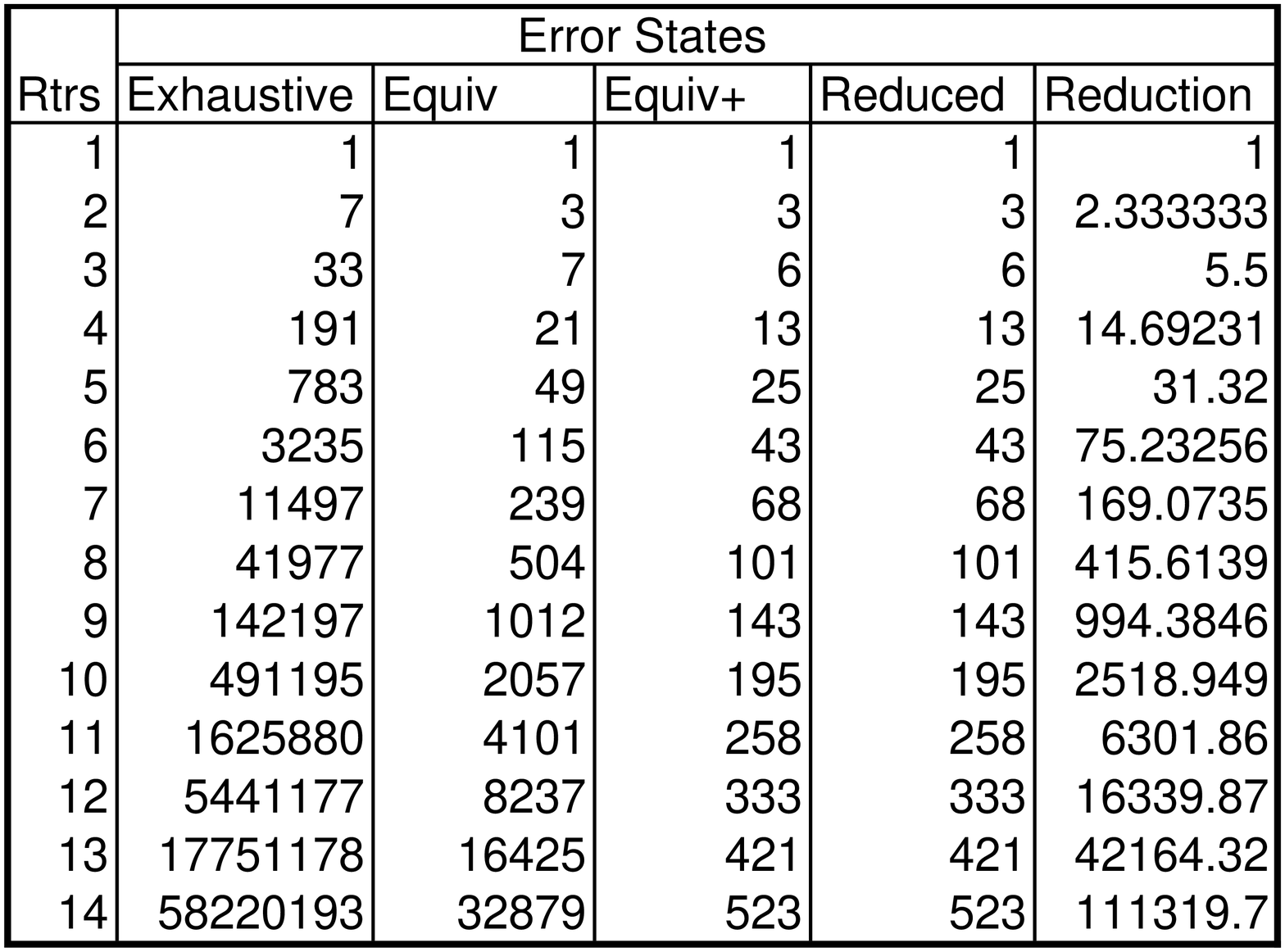,height=4cm,width=8cm,clip=,angle=0} 
   \caption{Simulation statistics for
forward algorithms. The number of stable error
states reached.}\label{tbl2.5}
 \end{center}
\end{figure}

\scriptsize

\par \hspace{0.05in}{\bf EquivInit}($S$,$i$,$GState$)\{
\par \hspace{0.1in}$\forall state \in S$
\par \hspace{0.3in}for $j = i$ to $0$ \{
\par \hspace{0.5in}$New$ = $emptystate$;   
\par \hspace{0.5in}for $k = 0$ to $j$      
\par \hspace{0.7in}add $state$ to $New$;   
\par \hspace{0.5in}$New = New \cdot GState$
\par \hspace{0.5in}$\bar{S} = $ trunc($S$,$state$);
\par \hspace{0.5in}if ($i-j$) = 0
\par \hspace{0.7in}ExpandSpace($New$);
\par \hspace{0.5in}else
\par \hspace{0.7in}EquivInit($\bar{S}$,$i-j$,$New$);
\par \hspace{0.3in}\}
\par \hspace{0.05in}\} \\

\normalsize

This procedure is invoked with the following parameters:
(a) the initial set of states $I.S.$ as $S$, (b) the number of
routers $n$ as $i$, and (c) the $emptystate$ as $GState$.
The procedure is recursive and produces the set of equivalent
initial states and invokes the ExpandSpace procedure for each
equivalent initial state.
The `trunc' function truncates $S$ such that $\bar{S}$ contains
only the state elements in $S$ after the element $state$. For
example, trunc($\{F,NM,M\},F$) = $\{NM,M\}$.

\item The second reduction we use is during state comparison.
Instead of comparing the actual states, we compare and store equivalent
states. Hence, the line `if $New \notin W$ or $V$' would check for
equivalent states. We call the algorithm after this second
reduction {\bf Equiv+}.

\item The third reduction is made to eliminate redundant transitions.   
To achieve this reduction we add flag check before invoking
{\bf forward}, such as stateFlag. The flag is set to 1 when the
stimuli for that specific state have been applied. We call the 
algorithm after the third reduction the {\bf reduced} algorithm.

\end{enumerate}

\subsection{Complexity analysis of forward search for PIM-DM}
\label{complexity_apndx}


The number of reachable states visited, the number of
transitions and the number of erroneous states found were
recorded. The result is given in
Figures~\ref{tbl1},~\ref{tbl1.5},~\ref{tbl2},~\ref{tbl2.5}.
The reduction is the ratio of the numbers obtained using the
exhaustive algorithm to those obtained using the {\bf reduced}
algorithm.

The number of expanded states denotes the number of visited  
stable states and is measured simply as the number of states in
the set $V$ in `ExpandSpace' procedure. The number of forwards is
the number of times the `forward' procedure was called denoting
the number of transitions between stable states. The number of
transitions is the number of visited transient states that are
increased with every new state visited in the `forward'
procedure.
The number of error states is the number of stable (or
expanded) states violating the correctness conditions. 

\begin{figure}[th]
 \begin{center}   
  \epsfig{file=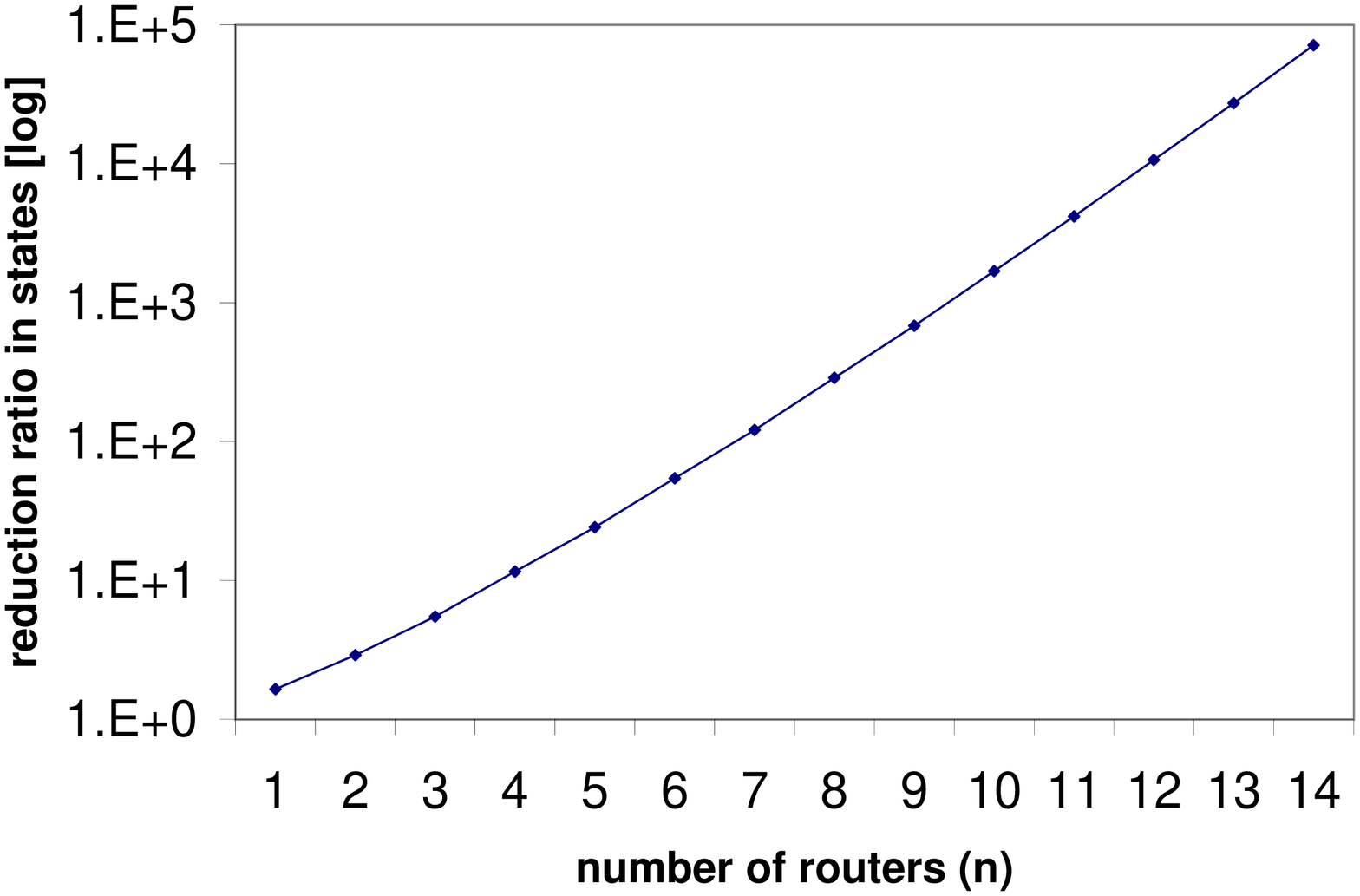,height=5cm,width=6cm,clip=,angle=0}
   \caption{Reduction ratio from exhaustive to 
the reduced algorithm}\label{reduction_fig}
 \end{center}
\end{figure}

The number of transitions is reduced from $O(4^n)$ for the
exhaustive algorithm 
to $O(n^4)$ for the {\bf reduced} algorithm.
This means that we have obtained exponential reduction in complexity, as
shown in Figure~\ref{reduction_fig}.

  

\section{FOTG Algorithms}
\label{fotg_apndx}


This appendix includes pseudo-code for procedures implementing the 
fault-oriented test generation (FOTG) method presented in 
Section~\ref{fotg}. In addition, it includes detailed results of our case 
study to apply FOTG to PIM-DM.

\subsection{Pre-Conditions}

The procedure described below takes as input the set of post-conditions for the FSM
stimuli and genrates the set of pre-conditions.
The `$conds$' array contains the post-conditions (i.e., the effects of
the stimuli on the system) and is indexed by the stimulus. The `stimulus'
function returns the stimulus (if any) of the condition. The `transition'
function returns the transition or state of the
condition~\footnote{If there's a state in the condition, this   
may be viewed as $state \rightarrow state$ transition, i.e.,
transition to the same state.}.
The pre-conditions are stored in an array `$preConds$' indexed
by the stimulus.

\scriptsize

\par \hspace{0.05in}{\bf PreConditions}\{
\par \hspace{0.1in}$\forall stim \in \tau$
\par \hspace{0.3in}$\forall cond \in conds[stim]$\{
\par \hspace{0.5in}$s$ = stimulus($cond$);
\par \hspace{0.5in}$t$ = transition($cond$);
\par \hspace{0.5in}add $t.stim$ to $preConds[s]$;
\par \hspace{0.3in}\}
\par \hspace{0.05in}\}

\normalsize

\subsection{Dependency Table}

The `dependencyTable' procedure generates the dependency table
$depTable$ from the transition table of conditions $conds$.

\scriptsize

\par \hspace{0.05in}{\bf dependencyTable}\{
\par \hspace{0.1in}$\forall stim \in \tau$
\par \hspace{0.3in}$\forall cond \in conds[stim]$ \{  
\par \hspace{0.5in}$endState$ = end($cond$);
\par \hspace{0.5in}$startState$ = start($cond$);
\par \hspace{0.5in}add $startState.stim$ to $depTable[endState]$;
\par \hspace{0.3in}\}
\par \hspace{0.05in}\}

\normalsize

For each state $s$, that is $endState$ of a transition, a set of
$startState$ -- $stimulus$ pairs leading to the creation of $s$ is
stored in the $depTable$ array.
For $s \in I.S.$ a symbol denoting initial state is added to the
array entry. For our case study $I.S. = \{NM,EU\}$.   

\subsection{Topology Synthesis}

The following procedure synthesizes minimum topologies necessary
to trigger the various stimuli of the protocol. It performs the
third and forth steps of the topology synthesis procedure explained in
Section~\ref{fotg_details}. 

\scriptsize

\par \hspace{0.05in}{\bf buildMinTopos}($stim$)\{
\par \hspace{0.1in}$\forall cond \in preConds[stim]$\{
\par \hspace{0.3in}$st$ = end($cond$);
\par \hspace{0.3in}$stm$ = stimulus($cond$);
\par \hspace{0.5in}if type($stm$) = $orig$
\par \hspace{0.7in}add $st$ to $MinTopos[stim]$;
\par \hspace{0.5in}else \{
\par \hspace{0.7in}if $\not\exists Topo(stm)$
\par \hspace{0.9in}buildMinTopos($stm$);
\par \hspace{0.7in}$\forall topo \in MinTopos[stim]$
\par \hspace{0.9in}add $st$ to $MinTopos[stim]$;
\par \hspace{0.5in}\}
\par \hspace{0.1in}\} 
\par \hspace{0.05in}\}

\normalsize

\subsection{Backward Search}

The `Backward' procedure calls the `Rewind' procedure to perform
the backward search. A set of visited states $V$ is kept to avoid
looping. For each state in $GState$ possible backward
implications are attempted to obtain valid backward steps toward
initial state. `Backward' is called recursively for preceding
states as a depth first search. If all backward branches are
exhausted and no initial state was reached the state is declared
unreachable.

\scriptsize

\par \hspace{0.05in}{\bf Backward}($GState$)\{
\par \hspace{0.1in}if $GState \in V$
\par \hspace{0.3in}return $loop$
\par \hspace{0.1in}add $GState$ to $V$
\par \hspace{0.1in}$\forall s \in GState$\{
\par \hspace{0.3in}$bkwds$ = $depTable[s]$;
\par \hspace{0.3in}$\forall bk \in bkwds$\{
\par \hspace{0.5in}$New$ = Rewind($bk$,$GState$,$s$);
\par \hspace{0.5in}if $New$ = done
\par \hspace{0.7in}break;
\par \hspace{0.5in}else
\par \hspace{0.7in}Backward($New$);
\par \hspace{0.3in}\}
\par \hspace{0.1in}\}
\par \hspace{0.1in}if all states are done
\par \hspace{0.3in}return reached
\par \hspace{0.1in}else
\par \hspace{0.3in}return unreachable
\par \hspace{0.05in}\}

\normalsize

The `Rewind' procedure takes the global state one step backward
by applying the reverse transition rules.
`replace($s$,$st$,$GState$)' replaces $s$ in $GState$ with $st$
and returns the new global state. Depending on the stimulus type
of the backward rule $bk$, different states in $GState$ are
rolled back. For $orig$ and $dst$ only the originator and
destination of the stimulus is rolled back, respectively.
For $mcast$, all affected states are rolled back except
the originator. $mcastDownstream$ is similar to $mcast$
except that all downstream routers or states are rolled back,
while only one upstream router (the destination) is rolled back.

\scriptsize

\par \hspace{0.05in}{\bf Rewind}($bk$,$GState$,$s$)\{
\par \hspace{0.1in}if $bk \in I.S.$
\par \hspace{0.3in}return done;
\par \hspace{0.1in}$stim$ = stimulus($bk$);
\par \hspace{0.1in}$st$ = start($bk$);
\par \hspace{0.1in}if type($stim$) = $orig$ \{
\par \hspace{0.3in}$New$ = replace($s$,$st$,$GState$);
\par \hspace{0.3in}return $New$;
\par \hspace{0.1in}\}
\par \hspace{0.1in}$\forall cond \in preconds[stim]$ \&
\par \hspace{0.1in}while $src$ not found \{
\par \hspace{0.3in}$str$ = start($cond$);
\par \hspace{0.3in}if $str \in GState$
\par \hspace{0.5in}$src$ found
\par \hspace{0.1in}\}
\par \hspace{0.1in}if $src$ not found
\par \hspace{0.3in}return backTrack;
\par \hspace{0.1in}if type($stim$) = $dst$ \{
\par \hspace{0.3in}$New$ = replace($s$,$st$,$GState$);   
\par \hspace{0.3in}if checkMinTopo($New$,$stim$)
\par \hspace{0.5in}return $New$;
\par \hspace{0.3in}else
\par \hspace{0.5in}return backTrack;
\par \hspace{0.1in}if not checkConsistency($stim$,$GState$)
\par \hspace{0.3in}return backTrack;
\par \hspace{0.1in}$New$ = $GState$;
\par \hspace{0.1in}if type($stim$) = $mcast$
\par \hspace{0.3in}$\forall cond \in conds[stim]$
\par \hspace{0.5in}if end($cond$) $\in GState$ \& not $src$
\par \hspace{0.7in}$New$ = replace(end,start,GState);
\par \hspace{0.1in}if type($stim$) = $mcastDownstream$
\par \hspace{0.3in}$\forall cond \in conds[stim]$
\par \hspace{0.5in}if end($cond$) $\in GState$ \& not $upstream$
\par \hspace{0.7in}$New$ = replace(end,start,GState);
\par \hspace{0.5in}else if $end \in GState$ \& $upstream$
\par \hspace{0.7in}$New$ = replace(end,start,GState) once;
\par \hspace{0.5in}if checkMinTopo($New$,$stim$)
\par \hspace{0.7in}return $New$;
\par \hspace{0.5in}else
\par \hspace{0.7in}return backTrack;
\par \hspace{0.05in}\}

\normalsize

\begin{figure}[t]
 \begin{center}
\vspace*{-.65cm}
\epsfig{file=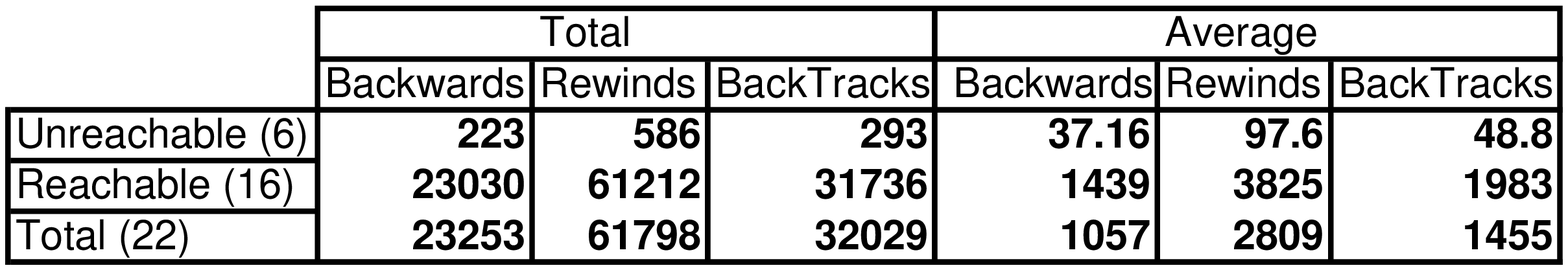,height=5cm,width=8cm,clip=,angle=0}
\vspace*{-.5cm}
   \caption{Case study statistics for applying FOTG to PIM-DM}\label{22topos}
\vspace*{-.5cm}
 \end{center}
\end{figure}

The following procedure checks for consistency of applying $stim$ to 
$GState$.

\scriptsize

\par \hspace{0.05in}{\bf checkConsistency}($stim$,$GState$)\{
\par \hspace{0.1in}$\forall cond \in conds[stim]$ \& $cond$ has transition
\par \hspace{0.3in}if start($cond$) $\in GState$
\par \hspace{0.5in}return False;
\par \hspace{0.3in}else
\par \hspace{0.5in}return True;
\par \hspace{0.05in}\}

\normalsize

The following procedure checks if $GState$ contains the necessary
components to trigger the stimulus.

\scriptsize

\par \hspace{0.05in}{\bf checkMinTopo}($GState$,$stim$)\{
\par \hspace{0.1in}if $\exists MinTopos[stim] \subseteq GState$
\par \hspace{0.3in}return  True;
\par \hspace{0.1in}else
\par \hspace{0.3in}return  False;
\par \hspace{0.05in}\}

\normalsize

\subsection{Simulation results}
\label{fotg_simulation}

We have conducted a case study of PIM-DM analysis using FOTG. A total of 22 topologies
were automatically constructed using as faults the selective loss of Join/Prune,
Graft, and Assert messages.
Out of the constructed topologies (or global states) 6 were unreachable global states
and 16 were reachable. The statistics for the total and average number of backward
calls, rewind calls and backtracks is given in Figure~\ref{22topos}.

Although the topology synthesis study we have presented above is not complete, we
have covered a large number of corner cases using only a manageable number of
topologies and search steps.

To obtain a complete representation of the topologies, we suggest to use the symbolic
representation~\footnote{We have used the repetition constructs `0', `1', `*'.}
presented in Section~\ref{search}.
Based on our initial estimates we expect the number of symbolic topology
representations to be approximately 224 topologies, ranging from 2 to 8-router LAN
topologies, for the single selective loss and single crash models.

\subsection{Experimental statistics for PIM-DM}

\begin{figure}[t]
 \begin{center}
\vspace*{-.65cm}
\epsfig{file=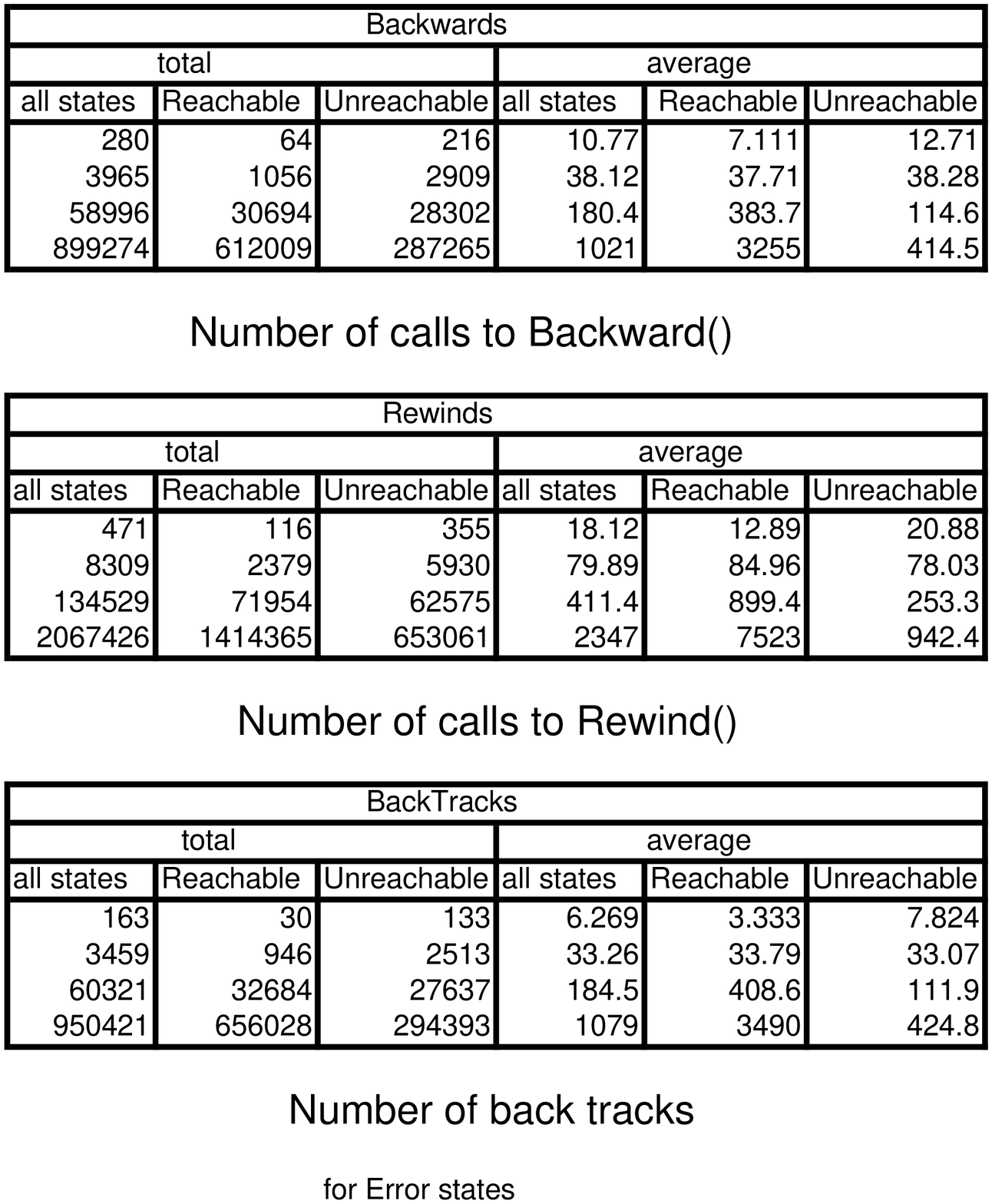,height=8cm,width=8cm,clip=,angle=0}
\vspace*{-.5cm}
   \caption{Simulation statistics for
backward algorithms}\label{bkwd2}
\vspace*{-.5cm}
 \end{center}
\end{figure}

To investigate the utility of FOTG as a verification tool we ran
this set of simulations. This is not, however, how FOTG is used to 
study protocol robustness (see previous section for case study analysis). 

We also wanted to study the effect of unreachable states on the 
complexity of the verification. The simulations for our case study show that 
unreachable states do not contribute in a significant manner to the 
complexity of the backward search for larger topologies. Hence, in order 
to use FOTG as a verification tool, it is not sufficient to add 
the reachability detection capability to FOTG.

The backward search was applied to the equivalent error states (for LANs with 2 to 5
routers connected).
The simulation setup involved a call to a procedure
similar to `EquivInit' in Appendix II-B, 
with the parameter $S$ as the set of state symbols, and after an error 
check was done a call is made to the `Backward' procedure instead of 
`ExpandSpace'.

\begin{figure}[th]
 \begin{center}
  \epsfig{file=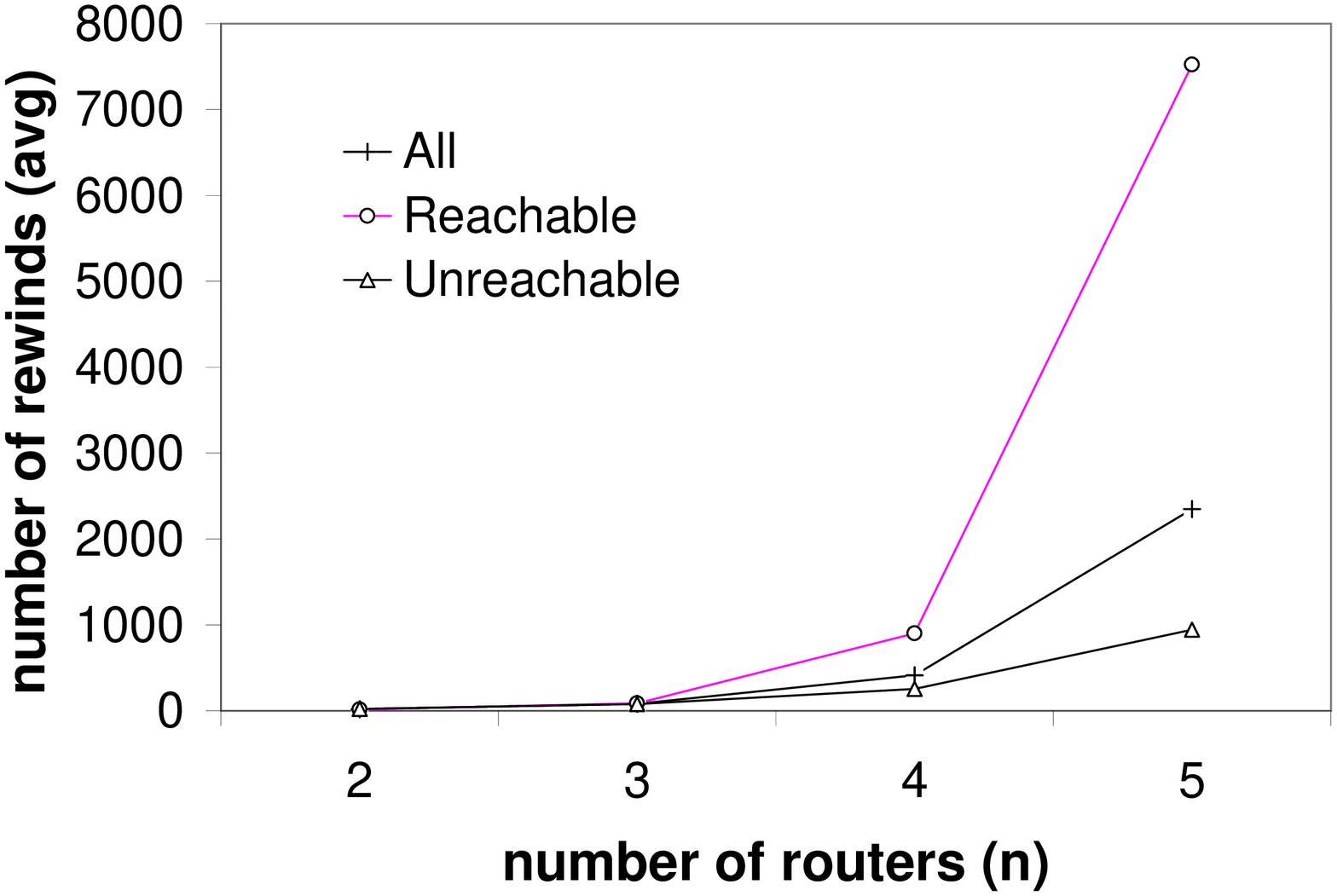,height=5cm,width=6cm,clip=,angle=0}
   \caption{Complexity of the FOTG algorithm for error
states}\label{fotg_complex_err}
 \end{center}
\end{figure}

\begin{figure}[th]
 \begin{center}
  \epsfig{file=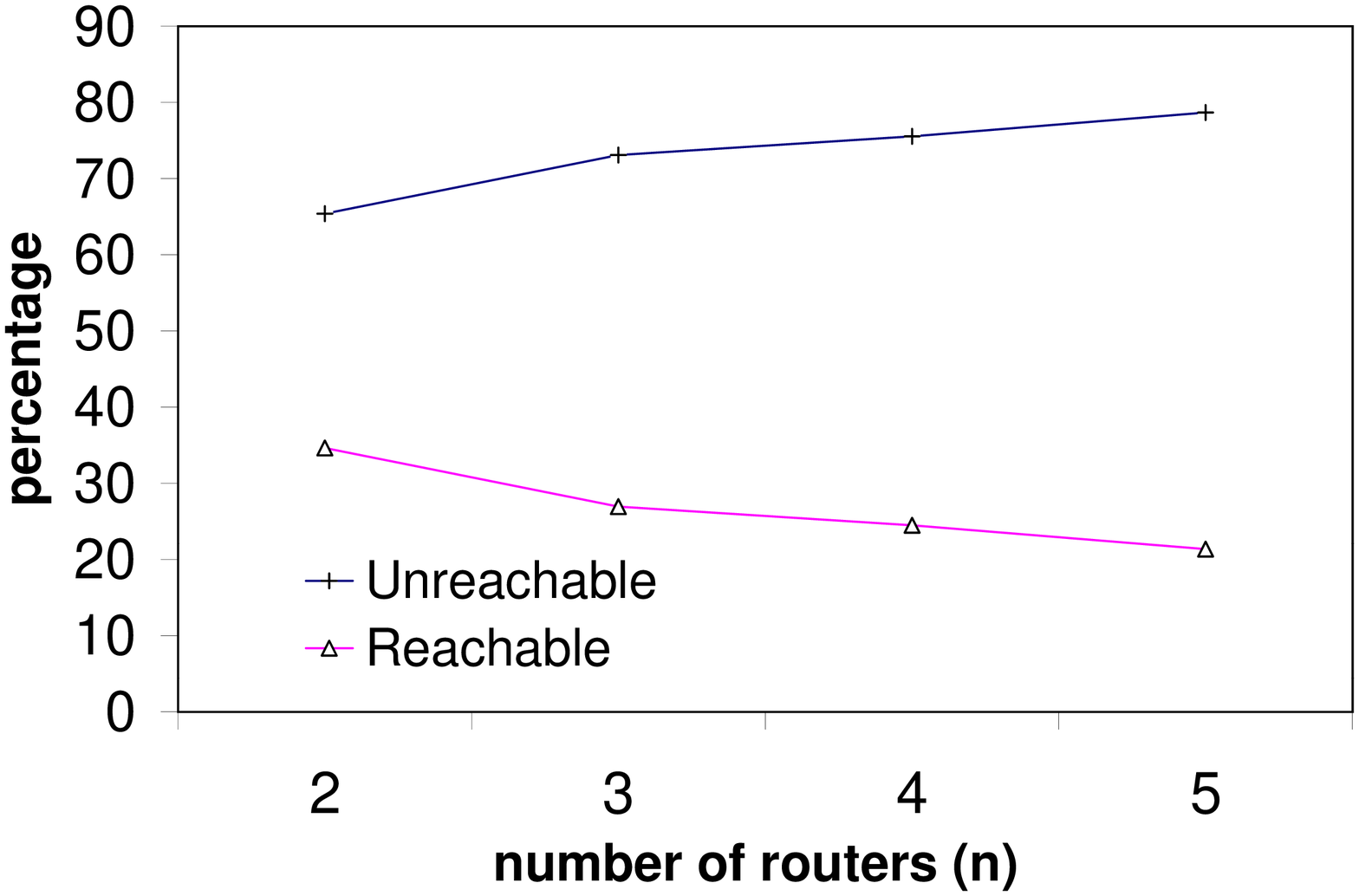,height=5cm,width=6cm,clip=,angle=0}
   \caption{Percentage of reachable/unreachable error states using
FOTG}\label{reach_perc_err}
 \end{center}
\end{figure}

States were classified as reachable or unreachable. 
For the four topologies
studied (LANs with 2 to 5 routers) statistics were measured
(e.g., max, min, median, average, and total) for number of calls
to the `Backward' and `Rewind' procedures, and the number of
backTracks were measured.
As shown in Figure~\ref{bkwd2}, the statistics
show that, as the topology grows, all 
the numbers for the reachable states get significantly larger
than those for the unreachable states (as in
Figure~\ref{fotg_complex_err}), despite the fact that
that the percentage of unreachable states increases with the topology
as in Figure~\ref{reach_perc_err}. 
The reason for such behavior is due to the fact that when the 
state is unreachable the algorithm reaches a {\em dead-end} relatively early 
(by exhausting one branch of the search tree).
However, for reachable states, the algorithm keeps on
searching until it reaches an initial global state.
Hence the reachable states search constitutes the major component
that contributes to the complexity of the algorithm.

\subsection{Results}
\label{detailed_results}

We have implemented an early version of the algorithm in the 
NS/VINT environment (see http://catarina.usc.edu/vint)
and used it to drive detailed simulations of PIM-DM therein, 
to verify our findings. 
In this section we discuss the results of applying our method to 
PIM-DM.
The analysis is conducted for single selective message loss.

For the following analyzed messages, we present the steps for
topology synthesis, forward and backward implication.

\subsubsection{Join} 
Following are the resulting steps for
$join$ loss:

\scriptsize

\vspace{.05in}
\hspace{-.6cm}
\begin{tabular}{|p{8cm}|} \hline
{\bf Synthesizing the Global State} \\ \hline
 1. Set the inspected message to $Join$ \\ \hline
 2. The $startState$ of the post-condition is $F_{dst\_Del}
\Longrightarrow
G_I = \{F_{j\_Del}\}$ \\ \hline
 3. The state of the pre-condition is $NH_i \Longrightarrow G_I
=
\{NH_i,F_{j\_Del}\}$ \\ \hline
 4. The stimulus of the pre-condition is $Prune$. Set the
inspected
message to $Prune$ \\ \hline
 5. The $startState$ of the post-condition is $F_j$ which can
be implied from $F_{j\_Del}$ in $G_I$ \\ \hline
 6. The state of the pre-condition is $NC_k \Longrightarrow
G_I = \{NH_i,F_{j\_Del}, NC_k\}$ \\ \hline
 7. The stimulus of the pre-condition is $L$. Set the inspected
message
to $L$ \\ \hline
 8. The $startState$ of the post-condition is $NH$ which can be
implied
from $NC$ in $G_I$ \\ \hline
 9. The state of the pre-condition is $Ext$, an external event
\\ \hline 
 {\bf Forward implication} \\ \hline
 without loss: $G_I = \{NH_i,F_{j\_Del},NC_k\}  
\stackrel{Join}{\longrightarrow}
G_{I+1} = \{NH_i,F_j,NC_k\}$ correct state \\ \hline
 loss w.r.t. $R_j$:
$\{NH_i,F_{j\_Del},NC_k\} \stackrel{Del}{\longrightarrow}
G_{I+1} = \{NH_i,NF_j,NC_k\}$ error state \\ \hline
 {\bf Backward implication} \\ \hline
 $G_I = \{NH_i,F_{j\_Del},NC_k\} 
\stackrel{Prune}{\longleftarrow} G_{I-1} =
\{NH_i,F_j,NC_k\} 
\stackrel{FPkt}{\longleftarrow} G_{I-2} = \{M_i,F_j,NM_k\}$
 $\stackrel{SPkt}{\longleftarrow}
G_{I-3} = \{M_i,EU_j,NM_k\}
\stackrel{HJ_i}{\longleftarrow} G_{I-4} = \{NM_i,EU_j,NM_k\} = I.S.$ \\ 
\hline \end{tabular}

\normalsize
\vspace*{.03in}
Losing the $Join$ by the forwarding router $R_j$ leads to an
error state 
where router $R_i$ is expecting packets from the LAN, but the LAN
has no 
forwarder.

\subsubsection{Assert}
Following are the resulting steps for the $Assert$ loss:

\scriptsize

\vspace{.05in}
\hspace{-.6cm}
\begin{tabular}{|l|} \hline
{\bf Synthesizing the Global State} \\
\hline
 1. Set the inspected message to $Assert$ \\ \hline
 2. The $startState$ of the post-condition is $F_j
\Longrightarrow G_I =
\{F_j\}$ \\ \hline
 3. The state of the pre-condition is $F_i \Longrightarrow G_I = 
\{F_i,F_j\}$ \\ \hline
 4. Stimulus of pre-condition is $FPkt_j$. Set
inspected
message to $FPkt_j$ \\ \hline
 5. The $startState$ of the post-condition is $EU_i$,
implied from $F_i$ in $G_i$ \\ \hline
 6. The state of the pre-condition is $F_j$, already in $G_I$ \\
\hline
 7. Stimulus of pre-condition is $SPkt_j$. Set
inspected
message to $SPkt_j$ \\ \hline
 8. The $startState$ of the post-condition is $NF_j$,
implied
from $F_j$ in $G_I$ \\ \hline
 9. The stimulus of the pre-condition is $Ext$, an external
event \\ 
\hline \hline
 {\bf Forward Implication} \\ \hline
 $G_I = \{F_i,F_j\} 
\stackrel{Assert_i}{\longrightarrow} G_{I+1} = \{F_i,NF_j\}$ error
\\ 
\hline \hline
 {\bf Backward Implication} \\ \hline
 $G_I = \{F_i,F_j\} 
\stackrel{FPkt_j}{\longleftarrow}
G_{I-1} = \{EU_i,F_j\}
\stackrel{SPkt_j}{\longleftarrow}
G_{I-2} = \{EU_i,EU_j\} = I.S.$ \\ \hline
\end{tabular}

\normalsize

\begin{figure}
 \begin{center}  
\vspace*{-.1in}
\epsfig{file=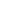,height=7cm,width=4cm,angle=270,clip=}
\vspace*{-.2in}
   \caption{A topology having a $\{F_i, F_j, \dots , F_k\}$  
LAN}\label{assert_topo}
\vspace*{-.15in}
 \end{center}
\end{figure}

\vspace*{.03in}
The error in the $Assert$ case occurs even in the absence of
message loss.
This error occurs due to the absence of a prune to stop the flow
of
packets to a LAN with no downstream receivers. This problem
occurs for
topologies with $G_I = \{F_i, F_j, \dots , F_k\}$, as that shown
in  
Figure~\ref{assert_topo}.

\subsubsection{Graft}
Following are the resulting steps for the $Graft$ 
loss:

\scriptsize

\vspace{.05in}
\hspace{-.6cm}
\begin{tabular}{|l|} \hline
{\bf Synthesizing the Global State} \\ \hline
 1. Set the inspected message to $Graft_{Rcv}$ \\ \hline
 2. The $startState$ of the post-condition is $NF
\Longrightarrow G_I = 
\{NF\}$ \\ \hline
 3. the $endState$ of the pre-condition is $NH_{Rtx}
\Longrightarrow G_I =
\{NF,NH_{Rtx}\}$ \\ \hline
 4. The stimulus of the pre-condition is $Graft_{Tx}$ \\ \hline
 5. The $startState$ of the post-condition is $NH$,
implied from $NH_{Rtx}$ in $G_I$ \\ \hline
 6. the $endState$ of the pre-condition is $NH$ which may be
implied \\ 
\hline
 7. the stimulus of the pre-condition is $HJ$, which is $Ext$
(external) \\ \hline \hline
 {\bf Forward Implication } \\ \hline
 without loss: $G_I = \{NH,NF\} 
\stackrel{Graft_{Tx}}{\longrightarrow}
G_{I+1} = 
\{NH_{Rtx},NF\}$ \\
 $
\stackrel{Graft_{Rcv}}{\longrightarrow}
G_{I+2} = \{NH_{Rtx},F\} 
\stackrel{GAck}{\longrightarrow} G_{I+3}
= \{NH,F\}$
correct state \\ \hline
with loss of $Graft$: $G_I = \{NH,NF\} 
\stackrel{Graft_{Tx}}{\longrightarrow}
G_{I+1} = \{NH_{Rtx},NF\} 
\stackrel{Timer}{\longrightarrow}$ \\
$G_{I+2} = \{NH,NF\} 
\stackrel{Graft_{Tx}}{\longrightarrow}
G_{I+3} =
\{NH_{Rtx},NF\} 
\stackrel{Graft_{Rcv}}{\longrightarrow}$ \\
$G_{I+4} = \{NH_{Rtx},F\}
\stackrel{GAck}{\longrightarrow}
G_{I+5} = \{NH,F\}$ correct state \\ \hline
\end{tabular}

\normalsize

\vspace*{.03in}
We did not reach an error state when the $Graft$ was lost, with
non-interleaving external events.


\subsection{Interleaving events and Sequencing}

A $Graft$ message is acknowledged by the $Graft-Ack$ ($GAck$)
message, and if not acknowledged it is retransmitted when the
retransmission timer expires. In an attempt to create an erroneous
scenario, the algorithm generates sequences to clear the retransmission
timer, and insert an adverse event. 
Since the $Graft$ reception causes an upstream router to become a
forwarder for the LAN, the algorithm interleaves a $Leave$ event as an
adversary event to cause that upstream router to become a non-forwarder.

To clear the retransmission timer, the algorithm
inserts the transition ($NH \stackrel{GAck}{\longleftarrow} NH_{Rtx}$)
in the event sequence.

{\bf Forward Implication}

$G_I = \{NH,NF\} 
\stackrel{Graft_{Tx}}{\longrightarrow}
G_{I+1} = \{NH_{Rtx},NF\}
\stackrel{GAck}{\longrightarrow}
G_{I+2} = \{NH,NF\}$ error state.

{\bf Backward Implication:}

Using backward implication, we can construct a sequence of events
leading to conditions sufficient to trigger the $GAck$. From the 
transition table these conditions are
$\{NH_{Rtx},F\}$\footnote{We do not 
show all branching or backtracking steps for simplicity.}:\\ 
$G_I = \{NH,NF\} 
\stackrel{HJ}{\longleftarrow}
G_{I-1} = \{NC,NF\} 
\stackrel{Del}{\longleftarrow}
G_{I-2} = \{NC,F_{Del}\} 
\stackrel{Prune}{\longleftarrow}
G_{I-3} = \{NC,F\} \stackrel{L}{\longleftarrow}
G_{I-4} = \{NH_{Rtx},F\}$.

To generate the $GAck$ we continue the backward implication 
and attempt to reach an initial state:\\
$G_{I-4} = \{NH_{Rtx},F\} 
\stackrel{Graft_{Rcv}}{\longleftarrow}
G_{I-5} =
\{NH_{Rtx},NF\} 
\stackrel{Graft_{Tx}}{\longleftarrow}
G_{I-6} = \{NH,NF\}  
\stackrel{HJ}{\longleftarrow}
G_{I-7} =
\{NC,NF\} 
\stackrel{Del}{\longleftarrow}
G_{I-8} = \{NC,F_{Del}\} 
\stackrel{Prune}{\longleftarrow}
G_{I-9} = \{NC,F\} 
\stackrel{FPkt}{\longleftarrow}
G_{I-10} = \{NM,F\} 
\stackrel{SPkt}{\longleftarrow}
G_{I-11} = \{NM,EU\} =
I.S.$

Hence, when a $Graft$ followed by a $Prune$ is 
interleaved with the $Graft$ loss, the retransmission timer is
reset with 
the receipt of the $GAck$ for the first $Graft$, and the systems
ends up in an error state.


\bibliographystyle{unsrt}
\bibliography{dissertation-2}

\end{document}